\documentclass[prb,twocolumn,preprintnumbers,amsmath,amssymb]{revtex4}
\usepackage{graphicx}% Include figure files
\usepackage{dcolumn}% Align table columns on decimal point
\usepackage{bm}% bold math

\usepackage[draft]{hyperref} %% optional
\usepackage{amssymb,amsmath}
\usepackage{multirow}
\usepackage{array}
\usepackage[T1]{fontenc}
\usepackage{textcomp}
\usepackage{ae}
\usepackage{soul} % for strikethrough
\usepackage{graphicx}
\usepackage{color}

\definecolor{BLUE}{rgb}{0,0,1} % tmp -- necessary to make the \wojtek macro in section titles work

\newcommand{\tsub}[1]{_{\text{#1}}}
\newcommand{\tXimin}{\tilde \Xi\tsub{min}}
\newcommand{\tXimax}{\tilde \Xi\tsub{max}}
\DeclareMathOperator{\interior}{int}
\DeclareMathOperator{\exterior}{ext}
\DeclareMathOperator{\RE}{Re}
\DeclareMathOperator{\IM}{Im}
\newcommand{\I}{\mathrm{i}}
\newcommand{\E}{\mathrm{e}}
\def\wojtek{\textcolor{blue}}          	% new text added by Wojtek in blue

\makeatletter
\renewcommand*\refstepcounter[1]{\stepcounter{#1}%
  \protected@edef\@currentlabel{%
    \csname p@#1\expandafter\endcsname
    \csname the#1\endcsname
  }%
}

\newcounter{arstructure}
\newcommand\declarestructure[1]{\refstepcounter{arstructure}\label{#1}}
\renewcommand\p@arstructure[1]{S#1}
\makeatother

\begin{document}

\title{Semi-analytical design of antireflection gratings for photonic crystals}

%% For REVTeX it is possible to automate superscript and e-mail callouts with the superscriptaddress option; see REVTeX4 documentation.

\author{Wojciech {\'S}migaj\\
\href{mailto:wojciech.smigaj@institut-optique.fr}{\texttt{wojciech.smigaj@institut-optique.fr}}}
\address{Laboratoire Charles Fabry, UMR 8501,
Institut d'Optique, CNRS, Universit\'e Paris Sud 11,
2 avenue Augustin Fresnel, 91127 Palaiseau Cedex, France.}

\author{Boris Gralak\\
\href{mailto:boris.gralak@fresnel.fr}{\texttt{boris.gralak@fresnel.fr}}}
\address{Institut Fresnel, CNRS, Aix-Marseille Universit\'e, Ecole Centrale Marseille,\\
Campus de St J\'er\^ome, 13397 Marseille Cedex 20, France.}

\begin{abstract}
This article concerns the design of antireflection structures which, placed on a photonic crystal
surface, significantly diminish the fraction of energy lost to reflected waves. After a review of
the classes of these structures proposed to date, a new method is presented in detail for the design
of antireflection gratings operating in a wide range of angles of incidence. The proposed algorithm
is illustrated by means of several examples, showing the advantages and limitations. 
\end{abstract}

%\pacs{42.70.Qs, % Photonic bandgap materials
%42.79.Wc, % Antireflection coatings
%42.25.Gy} % Edge and boundary effects; reflection and refraction

\maketitle %% null function with osajnl.sty

\declarestructure{ar:lens:t-14}
\declarestructure{ar:lens:t-0}
\declarestructure{ar:lens:coating-model-r0}
\declarestructure{ar:lens:coating-true-r0}
\declarestructure{ar:lens:binary-good}
\declarestructure{ar:lens:binary-bad}
\declarestructure{ar:lens:cr112}
\declarestructure{ar:lens:cr115}
\declarestructure{ar:lens:cr108}
\declarestructure{ar:lens:cr111}

\declarestructure{ar:coll:uncut-holes}
\declarestructure{ar:coll:cut-holes}
\declarestructure{ar:coll:coating-model-r0}
\declarestructure{ar:coll:coating-true-r0}
\declarestructure{ar:coll:grating-normal-pos}
\declarestructure{ar:coll:grating-shifted-pos}
\declarestructure{ar:coll:grating-optimised}

\declarestructure{ar:lens:flat-surf-modes}

\section{Introduction}

Photonic{-}crystal (PC) devices may become vital ingredients of integrated optical circuits. For
instance, they open new possibilities and functionalities like negative refraction
\cite{NotomiPRB00,PendryPRL00,GralakJOSAA00}, {supercollimation} \cite{EnochTAP03,GralakBook06},
compact isolators \cite{WangAPB05,SmigajOL10}, et c. On the other hand, PCs generally present
significant reflectivity at their surface while, in many applications, one strives after a perfect
energy transfer between the incident plane wave and the propagative PC eigenmode (or, more rarely,
several such modes). An important issue is then to eliminate propagative reflected waves. In this paper we propose a semi-analytical algorithm for the design of gratings acting as wide-angle antireflection (AR) structures for two-dimensional PCs.

The article is organized as follows. In the next section we review existing AR
structures for homogeneous media and PCs. Subsequently, in section
\ref{sec:ar:algorithm}, the three steps of the new design procedure are presented in detail. This algorithm is then tested on several particular cases: a PC flat lens (section
\ref{sec:ar:example-flat-lens}), a supercollimating PC (section
\ref{sec:ar:example-self-collimation}), and a non-reciprocal PC mirror (section
\ref{sec:ar:example-superprism}). Finally, in section \ref{sec:evanescent} we discuss briefly the influence of AR gratings in the evanescent-wave regime. 
% These examples are illustrated by numerical simulations performed
% with the methods which are briefly described in the appendix \ref{appendixB}.

\section{Background} \label{sec:ar:literature-review}

\begin{figure*}
  \centering
   \includegraphics{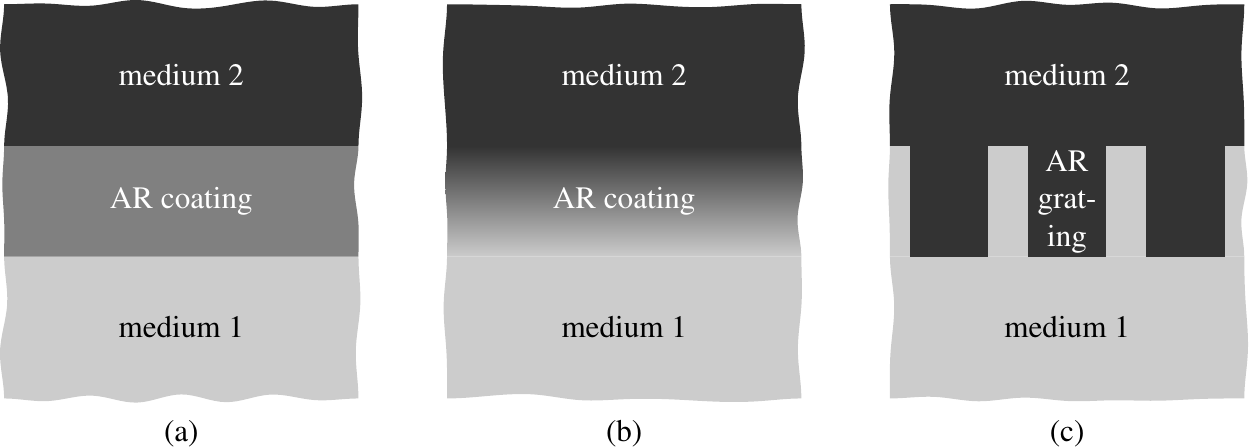}
\caption{Example AR structures belonging to the three principal classes described in the text.
Darker areas denote regions with higher refractive index. {(a)}~A homogeneous-layer AR
coating. {(b)}~An inhomogeneous-layer AR coating. {(c)}~A (binary lamellar) AR
grating.}
  \label{fig:ar:examples}
\end{figure*}

In this section we shall review existing methods of eliminating reflections from interfaces
which separate different media. We start with the simpler case of interfaces separating homogeneous
media, since the basic tools used to deal with this class of systems turn out to be valuable
also when more complex media, such as PCs, are involved. The existing AR solutions can be divided
into three broad classes. The refractive index~$n$ of \emph{AR coatings} depends
only on the coordinate~$z$ perpendicular to the interface; depending on whether $n(z)$ is
piecewise-constant or not, we speak of \emph{homogeneous-layer} or \emph{inhomogeneous-layer} AR
coatings \cite{DobrowolskiAO02}. In turn, the refractive index of \emph{AR gratings} is also a
function of the coordinates parallel to the interface. Figure \ref{fig:ar:examples} shows example AR
structures belonging to these three classes.

The simplest homogeneous-layer AR coating consists of a single thin film with refractive index~$n$
and thickness~$d$ chosen so as to ensure destructive interference of waves reflected from its top
and bottom surfaces, excited by a plane wave with a certain frequency~$\omega$ and angle of
incidence~$\theta$. The values of $n$ and~$d$ can be obtained analytically. In order to improve the
angular and frequency tolerance of the coating, one can increase the number of layers to make the
transition between the refractive index of the super- and substrate more gradual
\cite{DobrowolskiAO02}, thus reducing the amplitude of waves reflected on the individual
discontinuities of $n(z)$. In the limit of an infinite number of layers, one arrives at an
inhomogeneous-layer coating with a continuous monotonic profile $n(z)$. Several specific profiles
have been proposed in the literature (see ref.\ \onlinecite{GrannJOSAA96} for a review).
Unfortunately, these ``ideal'' AR coatings, even the single-layer one, cannot usually be realized
because of the lack of suitable materials with the required values of~$n$. This is the case, in
particular, for air-solid interfaces, where thin films with refractive index close to unity are
needed.

Short of using special ultra-low-index media, such as so-called Reststrahlen materials
\cite{DobrowolskiAO02}, there are two basic ways of circumventing this problem. One consists in
abandoning completely the structures based on a ``continuous transition'' between the super- and
substrate refractive indices in favor of interference-based coatings \cite{DobrowolskiAO02}. In
these systems, the total reflected wave vanishes thanks to the destructive interference of partial
waves generated at interfaces between layers with \emph{contrasting} values of~$n$. For instance,
so-called $v$-coatings consist of two layers with prescribed refractive indices $n_1$ and $n_2$
(corresponding to realistic materials) and thicknesses $d_1$ and $d_2$ adjusted so as to eliminate
reflection at the desired values of $\omega$ and~$\theta$. Note that $n_1$ and $n_2$ must satisfy
certain conditions in order that appropriate $d_1$ and $d_2$ can be found (ref.\ \onlinecite{MacleodBook01}, p.\
96). A disadvantage of $v$-type AR coatings is that their angular and frequency
tolerance are usually inferior even to those of the corresponding ideal single-layer coatings
{(ref.\ \onlinecite{MacleodBook01}, p.\ 97; ref.\ \onlinecite{OrfanidisBook08}, p.\ 188)}.

The other solution consists in using subwavelength gratings to simulate AR coatings with arbitrary
$n(z)$ profiles. In many cases, the effective-medium theory of gratings can be employed to calculate
the grating profile mimicking the desired $n(z)$ dependence \cite{GrannJOSAA96}. Several types of
gratings, such as the lamellar, trapezoidal, sinusoidal, triangular and pyramidal ones, have been
studied in the literature and shown to have good AR properties
\cite{GrannJOSAA96,BrauerAO94,RaguinAO93March,RaguinAO93May}. A review of the experimental methods
used to fabricate such AR structures can be found in ref.~\onlinecite{KikutaOR03}.

\begin{figure}
  \centering%\floatfont
  \includegraphics{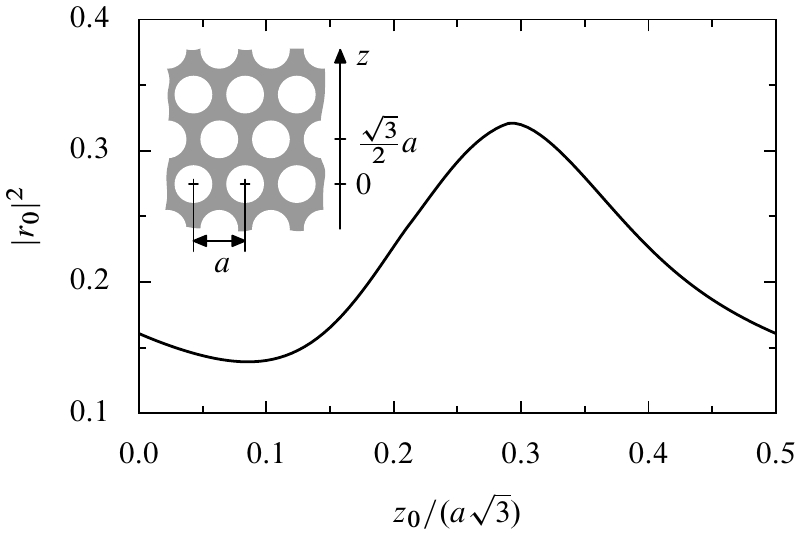}

  \caption{The dependence of the reflectance $|{r_0}|^2$ of the PC shown in the inset, placed in
    air, on the location of the truncation plane $z = z_0$. The PC consists of a hexagonal lattice
    of air holes with radius $0.365a$, where $a$ is the lattice constant, etched in a dielectric
    matrix with permittivity $\epsilon = 10.6$. The impinging wave is normally incident,
    $s$-polarized and has frequency $\omega = 0.311 \times 2\pi c/a$. It can be seen that
    $|{r_0}|^2$ does not fall under 0.13 for any truncation plane.}
  \label{fig:ar:dependence-on-trunc-planes}
\end{figure}

Let us now turn to the case of PCs.
One of their distinguishing features is the dependence of
their reflection coefficient on the position of their truncation plane. One could hope then that a
significant reduction of a PC's reflectance could be achieved without adding any AR structure, but
simply by choosing an appropriate cut. For some crystals, this has indeed proved to be possible
\cite{XiaoAPL04,BottenPRE06}. In particular, Botten \textit{et al.} \cite{BottenPRE06} have shown that very low
reflectance is a rather general feature of rod-type PCs truncated midway between successive layers
of rods. For many crystals, however, no truncation plane provides a sufficiently small value of
reflectance (see fig.\ \ref{fig:ar:dependence-on-trunc-planes}). This method of reducing reflection
is therefore not general enough, and one often has to resort to introducing some AR structure.
Several types of them have been proposed in the literature. More often than not, they have much in
common with one of the solutions developed with homogeneous materials in mind, reviewed in the
previous subsection.

The simplest approach has been proposed by Li \textit{et al.} \cite{LiJPD07}. It consists in placing in front of the
crystal a $v$-type AR coating composed of two layers made of constituent materials of the crystal,
usually air and a dielectric. Their thicknesses can be determined analytically or graphically as
soon as the reflection coefficient of the semi-infinite uncoated PC at the selected operation
frequency is known. This frequency is assumed to be low enough that only the zeroth diffraction
order be propagative in the layer closest to the PC surface, so that all higher orders reflected by
the crystal decay before reaching the boundary between the two layers. If it is not the case, the
structure may still work, but the optimum layer thicknesses will not in general be given by the
analytical formulas derived from the thin-film theory. The basic disadvantage of this otherwise very
attractive approach is the relatively low angular and frequency tolerance of $v$-type AR coatings.

Another group has studied a similar approach, in which one of the homogeneous layers is replaced by
a grating of rods or holes \cite{LeeOE08,ParkJOSAB09}. The performance of the resulting AR gratings
was shown to be similar to that of $v$-type coatings \cite{ParkJOSAB09}. Related structures, albeit
with only a single degree of freedom (the radius of the outermost rods of a rod-type PC), were also
analyzed by Jin and He \cite{JinPLA07}. In turn, Zhang and Li \cite{ZhangEPJD07} proposed a more complex wide-angle AR
grating for the interface between air and a PC flat lens, whose fabrication would be seriously
hindered, though, by the presence of extremely thin air slits and dielectric veins.

In contrast to the above works, which concentrated on the low-frequency regime, the authors of
refs.\ \onlinecite{BabaJJAP01,BabaOE04,MatsumotoOE05} endeavored to eliminate the reflection from an
interface between a PC and a semiconductor at a frequency where multiple propagative diffraction
orders existed in the latter medium. They proposed an AR grating consisting of elongated drop-shaped
air holes, this nonstandard form being motivated by the desire of ensuring a gradual transition
between the two media. In fact, the shape of the resulting structure resembles closely the classical
trapezoidal AR gratings studied, e.g., by Raguin and Morris \cite{RaguinAO93May}. The improvement brought about by
these gratings has subsequently been demonstrated experimentally \cite{MatsumotoOL06}.
Unfortunately, Baba et al.\ did not provide any analytical guidelines regarding the choice of the
geometrical parameters of the gratings, resorting instead to a time-consuming scan of the parameter
space in order to find the optimum structure \cite{BabaJJAP01,BabaOE04}.

A very important contribution was made by Lawrence \textit{et al.}\ \cite{LawrenceAPL08,LawrencePRA09}, who introduced the
concept of generalized \emph{matrix-valued} effective immittance~$\Xi$ of gratings and showed
that their effective-medium description can be made arbitrarily accurate by allowing~$\Xi$ to
have sufficiently large dimensions. Such matrix-valued immittance cannot serve as a drop-in
replacement of the corresponding scalar quantity in standard formulas derived for homogeneous media
\cite{LawrencePRA09}. Therefore, for instance, analytical determination of the optimum parameters of
an AR structure for a given PC composed of layers described by a matrix $\Xi$ is not possible.
However, Lawrence \textit{et al.}\ \cite{LawrencePRA09} derived equations similar to the classical Fresnel
formulas, but involving generalized immittances, 
and showed that accurate values of the scattering coefficients of typical PCs could be obtained already using immittance matrices truncated
to $5\times5$ elements or less. Since numerical calculations
involving such small matrices are very fast, it becomes feasible to design optimum AR gratings by
performing a full scan of the available parameter space. Example AR structures presented in
ref.~\onlinecite{LawrencePRA09} include, for instance, a relatively wide-band AR coating of a
superprism-type PC at a frequency where multiple propagative diffraction orders exist in the
adjacent dielectric. A slight limitation of the approach of Lawrence \textit{et al.}\ \cite{LawrencePRA09} is that the
generalized immittance has only been defined for 2D gratings symmetrical with respect to a two-fold
rotation axis parallel to the direction of invariance. Therefore, it does not cover, for instance,
triangular or trapezoidal gratings.

All the AR structures discussed so far are relatively compact, with thickness rarely exceeding one
or two lattice constants of the underlying PC. Some authors have advocated sacrificing compactness
in favor of potentially larger frequency and angular tolerance offered by thick stacks of gratings,
whose geometry changes gradually so as to ensure a smooth (``adiabatic'') transition of the
electromagnetic field of the incident plane wave towards the Bloch mode of the semi-infinite PC.
Several design principles for such gratings have been proposed \cite{WitzensPRE04,MomeniAPL05}. In
this work, we shall focus on compact AR structures, and therefore we omit a detailed discussion of adiabatic AR gratings.

\section{Design procedure}
\label{sec:ar:algorithm}
In this section we shall present an alternative method of designing AR gratings for PCs. Compared
to the approaches reviewed in section~\ref{sec:ar:literature-review}, it has the following
distinguishing features. First,
instead of performing a potentially time-consuming global scan of possible geometries, we use
an effective-medium model of PCs to calculate analytically the geometrical parameters of a
``tentative'' AR grating; then, if necessary, we refine these parameters with a numerical
local-minimiza\-tion algorithm. The final result is a trapezoidal AR grating. Second, the AR structures obtained in the proposed way have good angular tolerance, which in some important
applications of PCs is more significant than the frequency tolerance. For instance, the quality of
the image produced by a PC flat lens with effective refractive index $n \approx -1$ depends in the
first place on the angular range of incident plane waves which are transmitted through the lens with
little or no energy loss. On the other hand, frequency tolerance is not vital since $n$~can be close
to~$-1$ only in a narrow frequency band. In our developments we draw on the results of {Raguin and Morris} \cite{RaguinAO93May}, who demonstrated that triangular
and trapezoidal gratings allow a significant reduction of reflectance at interfaces between
homogeneous media in a wide range of incidence angles. 

\begin{figure}
  \centering%\floatfont
  \includegraphics{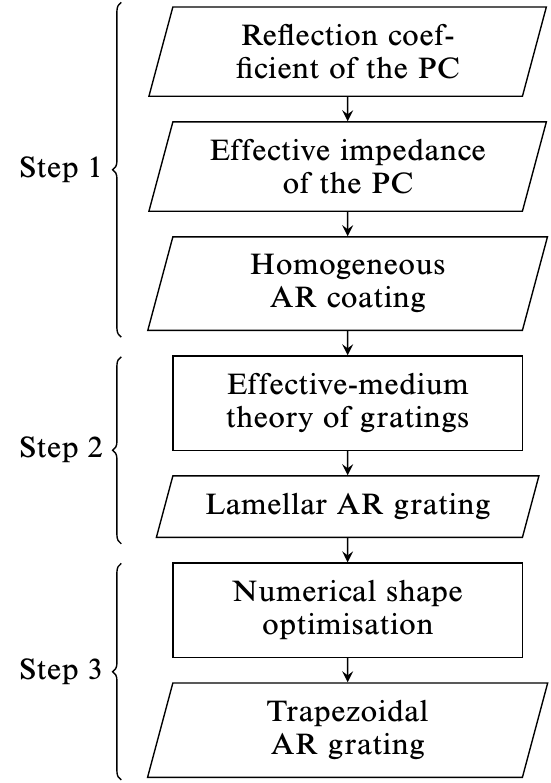}
  \caption{Successive steps of the proposed AR grating design algorithm.}
  \label{fig:ar:flowchart}
\end{figure}

The proposed design algorithm consists of three basic steps. In step~1 we use the value of the
reflection coefficient of the PC at a particular frequency and incidence angle to calculate
analytically the parameters of a homogeneous AR coating appropriate for the PC. In step~2 we convert
the coating into a lamellar AR grating composed of the same materials as the PC itself. If
necessary, in step~3 we adjust the shape of the grating using a numerical optimisation procedure,
obtaining finally a trapezoidal AR grating. We shall now proceed to detailing these three
constituent steps of the algorithm. They are summarised in the flowchart in fig.\
\ref{fig:ar:flowchart}.

\begin{figure}
  \centering%\floatfont
   \includegraphics{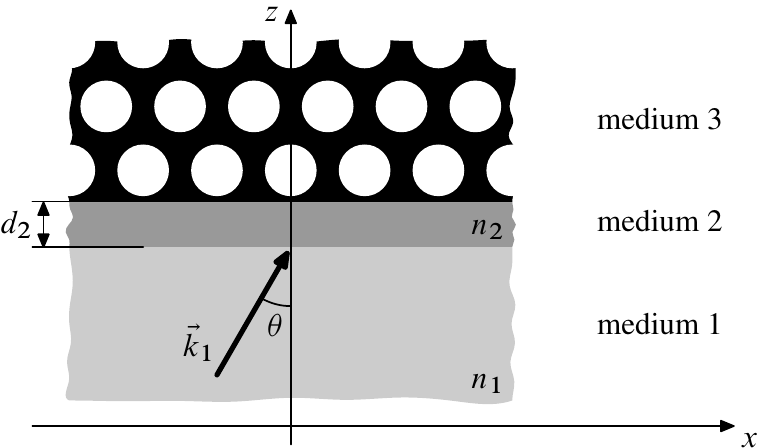}
  \caption{System considered in step~1 of the AR grating design procedure.}
  \label{fig:ar:system}
\end{figure}

\paragraph*{Step~1}
We begin by using the classical theory of AR coatings to calculate the
refractive index and thickness of a single-layer coating that should minimize the reflection from
the PC surface at a fixed angle of incidence~$\theta$ and frequency~$\omega$. The value of $\theta$
should lie approximately midway the desired angular operation range of the final AR structure.
Consider the system shown in fig.\ \ref{fig:ar:system}, in which a propagative plane wave with
frequency $\omega = c k_0$ and wave vector $\vec k_1 = (k_x, k_{z1}) = (n_1 k_0 \sin\theta, n_1 k_0
\cos\theta)$ impinges from the dielectric~1 with refractive index~$n_1$ on the surface of the
homogeneous film~2 with thickness~$d_2$ and refractive index~$n_2$ coating the PC~3. Let us assume
that medium~2 supports exactly one propagative diffraction order and is sufficiently thick for all
the evanescent orders reflected from the PC surface to vanish at the interface between media 1
and~2. We can calculate the total amplitude of the reflected plane wave, $r$, by summing up all the
multiple reflections occurring in the system:
\begin{equation}
  \label{eq:ar:multiple-refl}
  \begin{split}
  r &= r_{12} + t_{12} \Phi r_{23} \Phi
  \Big[\sum_{n=0}^\infty (r_{21} \Phi r_{23} \Phi)^n\Big] t_{21} \\
  &=
  \frac{r_{12} + (t_{12} t_{21} - r_{12} r_{21}) r_{23} \Phi^2}
  {1 - r_{21}r_{23}\Phi^2},
\end{split}
\end{equation}
where $r_{ij}$ and $t_{ij}$ denote the amplitudes of the waves reflected from the interface between
media $i$ and~$j$ and transmitted through it, respectively, and $\Phi \equiv {\E}^{{\I} k_{z2} d_2}$
with $k_{z2} \equiv (n_2^2 k_0^2 - k_x^2)^{1/2}$. The coefficients $r_{ij}$ and $t_{ij}$ are given by the Fresnel's formulas,
 \begin{equation}
  \label{eq:ar:fresnel-formulas}
  r_{ij} = \frac{\Xi_j - \Xi_i}{\Xi_j + \Xi_i},
  \qquad
  t_{ij} = \frac{2\Xi_j}{\Xi_j + \Xi_i},
\end{equation}
with $\Xi_i$ denoting the transverse immitance of medium~$i$ normalised to the immitance of free
space. For $s$~polarisation $\Xi_i$ is defined as the transverse impedance {of medium~$i$}: $\Xi_i \equiv Z_i
\equiv k_0 \mu_i/k_{z,i}$. {For} $p$~polarisation, it is defined as the transverse
admittance {of the medium}: $\Xi_i \equiv Y_i \equiv k_0 \epsilon_i/k_{z,i}$.
From eqs.\ \eqref{eq:ar:fresnel-formulas} it immediately follows that $r_{21} = -r_{12}$ and $t_{12} t_{21} - r_{12} r_{21} = 1$, hence
\begin{equation}
  \label{eq:ar:multiple-refl-simplified}
  r = \frac{r_{12} + r_{23} \Phi^2}{1 + r_{12}r_{23}\Phi^2}.
\end{equation}
The parameters of the antireflection coating, $n_2$ and~$d_2$, can now be obtained by requiring the
numerator of the fraction in the above equation to vanish. If the coating is lossless, so that
$|\Phi| = 1$, the numerator vanishes if and only if (i) the moduli of $r_{12}$ and $r_{23}$ are
equal and (ii) the thickness~$d_2$ is such that
\begin{equation}
  \label{eq:ar:condition-for-d-2}
  \arg r_{23} + 2k_{z2} d_2 = \arg r_{12} + (2m + 1)\pi,
\end{equation}
where $m$ is an integer and $\arg z$ stands for the argument of the complex number $z$. Solving
for~$d_2$, we get
\begin{equation}
  \label{eq:ar:d-2}
  d_2 = \frac{\arg r_{12} - \arg r_{23} + (2m+1)\pi}{2k_{z2}}.
\end{equation}
It is usually best to choose the value of $m$ corresponding to the smallest positive admissible
value of $d_2$; otherwise, internal resonances in the coating layer can spoil its antireflective
properties for some angles of incidence.

We shall now use condition (i) to determine the refractive index $n_2$ of the AR coating.
Substituting the Fresnel's formulas~\eqref{eq:ar:fresnel-formulas} into the condition
$|{r_{12}}|^2 = |{r_{23}}|^2$, we obtain
\begin{equation}
  \label{eq:ar:necessary-cond-explicit}
  \frac{(\Xi_2-\Xi_1)^2}{(\Xi_2+\Xi_1)^2} =
  \frac{(\Xi_3-\Xi_2)(\Xi_3^*-\Xi_2)}
  {(\Xi_3+\Xi_2)(\Xi_3^*+\Xi_2)}.
\end{equation}
Straightforward algebra leads to
\begin{equation}
  \label{eq:ar:necessary-cond-simplified}
  \Xi_2^2 = \frac{-\RE \Xi_3 + \left|{\Xi_3}\right|^2/\Xi_1}
  {\RE \Xi_3 - \Xi_1} \Xi_1^2.
\end{equation}
The $s$- and $p$-polarization cases need now to be considered separately. Assuming materials 1 and~2
to be nonmagnetic ($\mu_1 = \mu_2 = 1$), in the $s$-polarization case we have
\begin{equation}
  \label{eq:ar:s:Xi-2}
  \Xi_2 = Z_2 = \frac{k_0}{k_{z2}} = 
  \frac{k_0}{\sqrt{n_2^2 k_0^2 - n_1^2 k_0^2 \sin^2 \theta}},
\end{equation}
hence
\begin{equation}
  \label{eq:ar:s:n-2}
  n_2^2 = 
  n_1^2\sin^2\theta + \frac{1}{Z_2^2}.
\end{equation}
It is easily seen that $Z_2^2$ must be non-negative in order that $k_{z2}$ be real, as we have
assumed. For $p$~polarization,
\begin{equation}
  \label{eq:ar:p:Xi-2}
  \Xi_2 = Y_2 = \frac{n_2^2 k_0}{k_{z2}} = 
  \frac{n_2^2 k_0}{\sqrt{n_2^2 k_0^2 - n_1^2 k_0^2 \sin^2 \theta}}.
\end{equation}
This leads to the quadratic equation for $n_2^2$,
\begin{equation}
  \label{eq:ar:p:n-2-quadratic}
  n_2^4 - Y_2^2 n_2^2 + Y_2^2 n_1^2 \sin^2 \theta = 0,
\end{equation}
which has real solutions
\begin{equation}
  \label{eq:ar:p:n-2}
  n_2^2 = \frac12
  \biggl(Y_2^2 \pm \sqrt{Y_2^4 - 4 Y_2^2 n_1^2 \sin^2 \theta}\biggr)
\end{equation}
provided that $Y_2^4 - 4 Y_2^2 n_1^2 \sin^2 \theta \geq 0$. It can be shown that this condition,
together with the condition of real-valuedness of $k_{z2}$ [for \emph{both} solutions of
eq.~\eqref{eq:ar:p:n-2}], is fulfilled if and only if 
\begin{equation}
  \label{eq:ar:p:constraint-on-Y-2}
  Y_2^2 \geq 4 n_1^2 \sin^2 \theta.
\end{equation}

In practice, there are further constraints on the choice of the constituent material of the coating.
Other experimental issues aside, $n_2$ is bounded from below by the refractive index of air, and
from above, by the index at which a second propagative diffraction order appears at the given value
of $k_x$. In appendix \ref{appendixA} it is shown how such constraints of the general form
\begin{equation}
  \label{eq:ar:constraints-on-n}
  n\tsub{min}^2 \leq n_2^2 \leq n\tsub{max}^2
\end{equation}
can be transformed into equivalent constraints on the immittance of the PC, $\Xi_3$. {Let us denote by $\tilde\Xi_{{\alpha}}$ ($\alpha = 3$, $\min$, $\max$) the \emph{reduced} immittances $\tilde X_{{\alpha}}/\tilde X_1$.} It follows {then} that
\eqref{eq:ar:constraints-on-n} {holds if and only if} one of the two following {sets} of conditions on~$\tilde\Xi_3$ is
satisfied: 
\begin{equation}
  \label{eq:ar:constraints-on-eta-t-3-set-theory}
  \left\{
  \begin{array}{l}
     \tilde\Xi_3 \notin  \interior P_-  \\
    \tilde\Xi_3 \notin \interior C\tsub{min} \\
    \tilde\Xi_3 \notin  \exterior C\tsub{max}
  \end{array}
  \right.
\quad, \quad \mbox{ or } \quad
  \left\{
  \begin{array}{l}
     \tilde\Xi_3 \notin \exterior P_- \\ 
    \tilde\Xi_3 \notin \exterior C\tsub{min} \\
    \tilde\Xi_3 \notin  \interior C\tsub{max}
  \end{array}
  \right. \quad ,
\end{equation}
where $P_-$ stands for the half-plane $\RE \tilde \Xi_3 < 1$ {and $C_\alpha$ ($\alpha = \min$, $\max$) are the circles of radius $\frac{1}{2} |{1 - \tilde\Xi_\alpha^2}|$ centered at $
\bigl(\frac{1}{2}(1 + \tilde\Xi_\alpha^2), 0\bigr)$. The symbols $\interior A$ and $\exterior A$ denote the interior and exterior of a region~$A$.
To illustrate various possible geometries of the region of the complex $\tilde\Xi_3$~plane
determined by the constraints~\eqref{eq:ar:constraints-on-n} transformed into the
form~\eqref{eq:ar:constraints-on-eta-t-3-set-theory}, fig.\ \ref{fig:ar:sample-allowed-region} shows
the shape of this region for $s$~polarization and three distinct choices of the parameters
$n\tsub{min}$, $n\tsub{max}$, $n_1$ and~$\theta$.

\begin{figure*}
  \centering
   \includegraphics{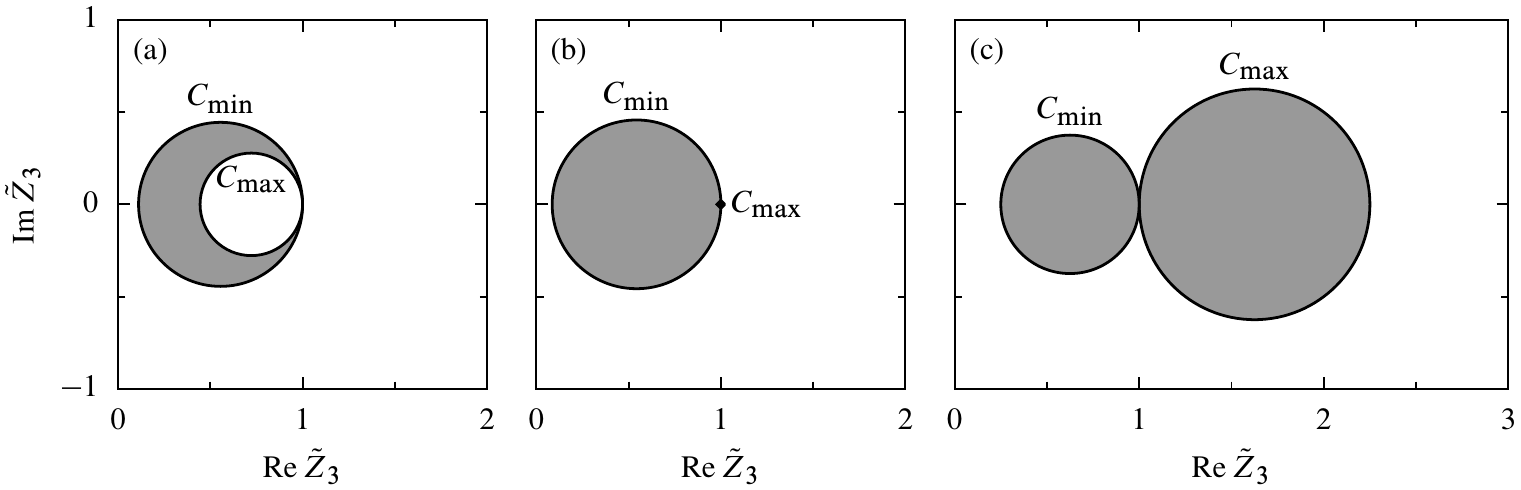}
\caption{{Shaded: r}egions of the complex $\tilde Z_3$ plane determined by the condition
\eqref{eq:ar:constraints-on-eta-t-3-set-theory} equivalent to the constraint
\eqref{eq:ar:constraints-on-n} for $s$~polarization and {(a)} $n_1 = 1$, $n\tsub{min} =
1.5$, $n\tsub{max} = 3$, $\theta = 0$, {(b)} $n_1 = 1$, $n\tsub{min} = 1$, $n\tsub{max} =
3$, $\theta = 30^\circ$, {(c)} $n_1 = 1.5$, $n\tsub{min} = 1$, $n\tsub{max} = 3$, $\theta
= 0$. The circles $C\tsub{min}$ and $C\tsub{max}$ are defined in the text after
eq.~\eqref{eq:ar:constraints-on-eta-t-3-set-theory}; note that in the case (b) $C\tsub{max}$
degenerates into the point $(1, 0)$.}
  \label{fig:ar:sample-allowed-region}
\end{figure*}

\paragraph*{{Step~2}} 
The coating obtained in step~1 is not practical, since its fabrication would
call for integration of the PC with a completely different solid; moreover, a suitable material with
the required value of refractive index might not be easily available. However, as noted in
section~\ref{sec:ar:literature-review}, a homogeneous thin film can often be replaced without
adverse effects by a subwavelength grating: this is {step 2}. Such a grating could be easily etched in the same process as the underlying PC; it would then naturally be composed of the same materials as the PC,
with permittivities, say, $\epsilon\tsub{l}$ and $\epsilon\tsub{h}$ ($\epsilon\tsub{l} < \epsilon\tsub{h}$).

In order to calculate the fill factor of a binary lamellar grating mimicking a layer with refractive
index~$n_2$ obtained in the previous step, one can resort to the classical second-order
effective-medium theory of gratings due to Rytov, described in ref.\ \onlinecite{RaguinAO93May}.
According to this theory, the effective permittivity~$\tilde\epsilon_s$ of a binary grating with
period~$a$ and fill factor~$f$ ($0 \leq f \leq 1$), composed of materials with permittivities
$\epsilon\tsub{l}$ and $\epsilon\tsub{h}$, and operating in the $s$~polarization is
\begin{equation}
  \label{eq:ar:s:eff-eps-2}
  \tilde\epsilon_s = \bar\epsilon_s
  \biggl[1 + 
  \frac{\pi^2}{3} \biggl(\frac{k_0a}{2\pi}\biggr)^2 f^2(1-f)^2 
  \frac{(\epsilon\tsub{h} - \epsilon\tsub{l})^2}
  {\bar\epsilon_s}\biggr],
\end{equation}
where
\begin{equation}
  \label{eq:ar:s:eff-eps-0}
  \bar\epsilon_s = f \epsilon\tsub{h} + (1-f) \epsilon\tsub{l}.
\end{equation}
For $p$~polarization, the effective permittivity $\tilde \epsilon_p$ is
\begin{equation}
  \label{eq:ar:p:eff-eps-2}
  \tilde\epsilon_p = \bar\epsilon_p
  \biggl[1 + 
  \frac{\pi^2}{3} \biggl(\frac{k_0a}{2\pi}\biggr)^2 f^2(1-f)^2 
  (\epsilon\tsub{h} - \epsilon\tsub{l})^2 
  \bar\epsilon_s
  \biggl(
  \frac{\bar\epsilon_p}{\epsilon\tsub{h} \epsilon\tsub{l}}
  \biggr)^2\biggr],
\end{equation}
where
\begin{equation}
  \label{eq:ar:p:eff-eps-0}
  \bar\epsilon_p = \biggl[\frac{f}{\epsilon\tsub{h}} + 
  \frac{1-f}{\epsilon\tsub{l}}\biggr]^{-1}
\end{equation}
and $\bar\epsilon_s$ is given by eq.~\eqref{eq:ar:s:eff-eps-0}.
Thus, the required fill factor can be obtained by setting $\tilde\epsilon_s$ or $\tilde\epsilon_p$
to $n_2^2$ in eq.\ \eqref{eq:ar:s:eff-eps-2} or \eqref{eq:ar:p:eff-eps-2} and solving it numerically
for~$f$.

It should be noted that in the domain of validity of Rytov's theory (small $k_0a/2\pi$) the
functions $\tilde\epsilon_s(f)$ and $\tilde\epsilon_p(f)$ are monotonically increasing from
$\epsilon\tsub{l}$ to $\epsilon\tsub{h}$. Thus, a binary grating cannot simulate a material with permittivity
outside the range delimited by the permittivities of the grating's constituent materials. As a
result, the bounds $n\tsub{min}^2$ and $n\tsub{max}^2$ mentioned in step~1 must fulfill 
$n\tsub{min}^2 \geq \epsilon\tsub{l}$ and $n\tsub{max}^2 \leq \epsilon\tsub{h}$, respectively.

\paragraph*{{Step~3}} 
The structure obtained at this stage should, in principle, ensure low
reflectance for incidence angles close to~$\theta$. Nevertheless, owing to the applied
approximations---neglect of higher diffraction orders excited by the PC and the AR grating---its
geometrical parameters might not be precisely optimal. In addition, it is well known
\cite{RaguinAO93May,RaguinAO93March} that trapezoidal and triangular AR gratings have larger angular
and frequency tolerance than lamellar ones. Therefore, as step 3, it is advisable to apply a
numerical optimization procedure to adjust the geometry of the grating, described by some small
number of parameters, so as to minimize a given objective function~$\rho$. The geometry obtained in
step~2 can be expected to provide a good starting point for a local search algorithm, such as the
Nelder-Mead simplex method {(ref.\ \onlinecite{NumericalRecipesInC}, section 10.4)}. 

\section{Application to a photonic crystal flat lens}

\label{sec:ar:example-flat-lens}

\begin{figure}
  \centering
   \includegraphics{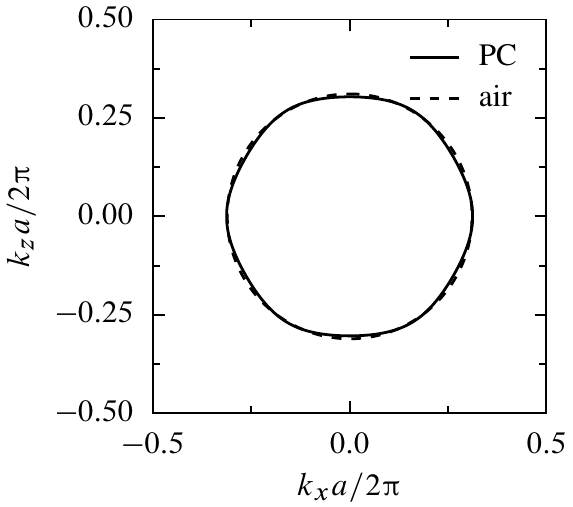}
\caption{Solid line: EFC of the PC considered in section~\ref{sec:ar:example-flat-lens} at frequency
$\omega = 0.311 \times 2\pi c/a$ and for $s$~polarization. Dashed line: EFC of air at the same
frequency.}
  \label{fig:ar:lens:efc}
\end{figure}

The first PC we shall consider is a hexagonal lattice of air holes of radius $r = 0.365a$, where $a$
is the lattice constant, etched in a dielectric matrix of permittivity $\epsilon = 10.6$. These
parameters correspond to the structure whose fabrication was reported in {ref.~\onlinecite{FabrePRL08}}. For $s$
polarization, at frequency $\omega = 0.311 \times 2\pi c/a$, the effective refractive index $n = -1$
can be attributed to the crystal, since its {equifrequency contour (EFC)} takes an approximately circular shape (fig.\
\ref{fig:ar:lens:efc}) with radius $K \approx \omega/c$ and group velocity directed inwards. {(Information about the techniques used to obtain the numerical results discussed in this paper can be found in appendix~\ref{appendixB}.)}

Veselago \cite{VeselagoSPU68} predicted that a slab of material with $n = -1$ should act as a flat lens: an image of an object placed near one of the surfaces of the slab should be produced on the other side
of the slab. Figure \ref{fig:ar:lens:field-maps-uncoated} shows the map of the modulus of the
electric field generated by a wire source with current 1\,A located above a slab of the PC in
question. The parts (a) and (b) refer to slabs truncated in
the ways shown in figs.\ \ref{fig:ar:lens:uncoated}(a) and~(b), respectively; from now on, these two
structures will be referred to as \ref{ar:lens:t-14} and \ref{ar:lens:t-0}. In accordance with the
theoretical predictions, images are formed below the slabs. However, their amplitude is low (67 and
79\,V$/$m for structures \ref{ar:lens:t-14} and \ref{ar:lens:t-0}, respectively) and intense beams
reflected from the top of the lenses are visible in the upper part of the plots. This suggests that
only a small fraction of energy is transmitted through the lenses. Indeed, as shown in fig.\
\ref{fig:ar:lens:uncoated}(a), the reflectance of structure~\ref{ar:lens:t-14},
$|{r_0(\theta)}|^2$, where $r_0$ is the specular reflection coefficient, exceeds 29\% for all
angles of incidence. Structure~\ref{ar:lens:t-0} performs better for low incidence angles, but
degrades quickly with increasing~$\theta$. We shall now apply the algorithm presented in
section~\ref{sec:ar:algorithm} to design an AR grating for this PC{, which ideally should function regardless of the angle of incidence.}

\begin{figure}
  \centering%\floatfont
   \includegraphics{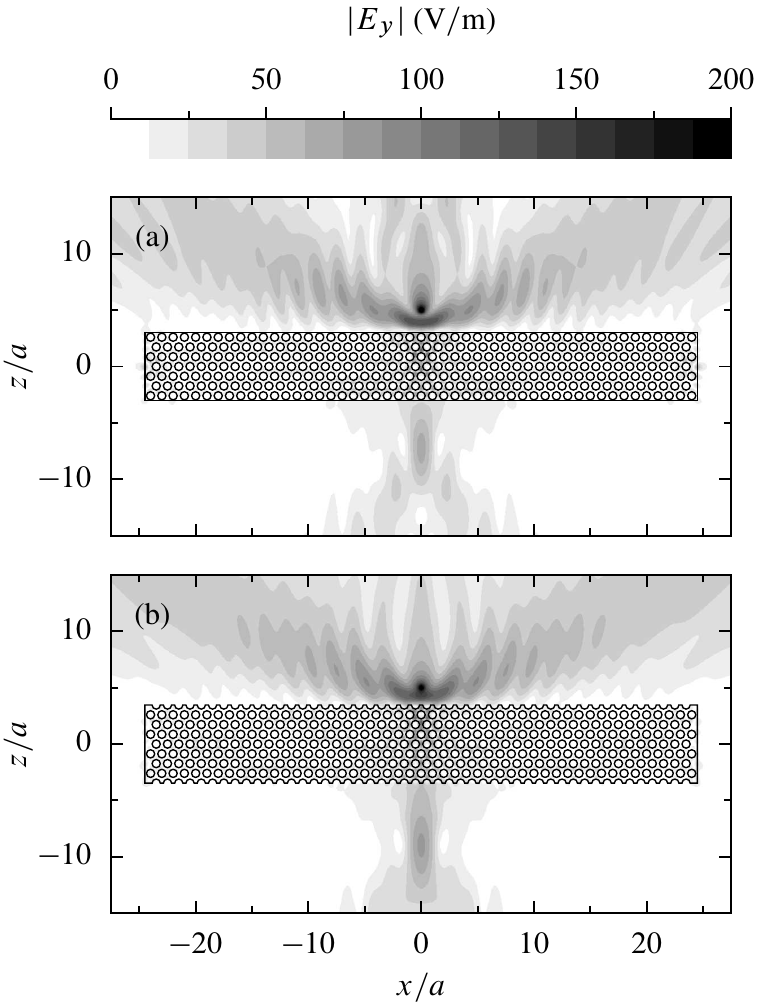}
 
\caption{Modulus of the electric field generated by an $s$-polarized wire source with current 1\,A
located above a slab of the PC studied in section~\ref{sec:ar:example-flat-lens} truncated along a
plane {(a)} lying midway between two neighboring rows of holes, {(b)} crossing
the centers of holes.}
  \label{fig:ar:lens:field-maps-uncoated}
\end{figure}

\begin{figure}
  \centering
  \includegraphics{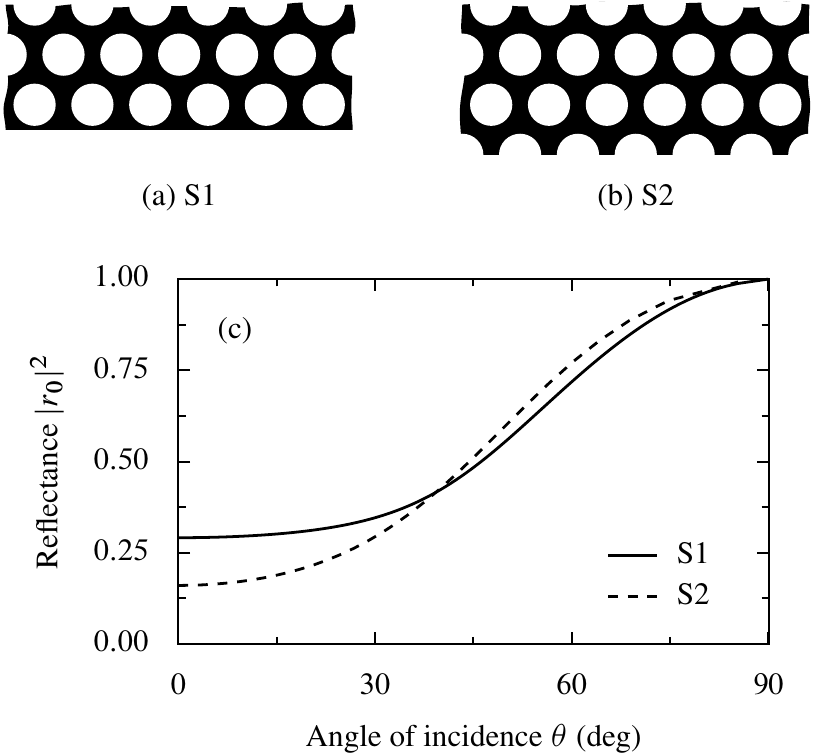}
  
\caption{Geometry of the PC studied in section~\ref{sec:ar:example-flat-lens} truncated along a
plane {(a)} lying midway between two neighboring rows of holes, {(b)} crossing
the centers of holes. {(c)}~Angular dependence of the reflectance of the structures shown
in parts (a) and (b).}
  \label{fig:ar:lens:uncoated}
\end{figure}

{As discussed in refs.\ \onlinecite{SmigajPRB08} and \onlinecite{SmigajSPIE08},} the
effective-medium model of PCs tends to be more accurate for crystals truncated along a plane with
constant permittivity profile, as is the case, for instance, for structure~\ref{ar:lens:t-14}.
Therefore in the first step of the design procedure we shall calculate the effective transverse
impedance~$Z_3$ of this structure. We consider two ways of obtaining this quantity. First, we
calculate it in the framework of the model presented in {ref.\ \onlinecite{SmigajPRB08}} using eq.~(18)
derived there. At frequency $\omega = 0.311 \times 2\pi c/a$ and angle of incidence $\theta =
45^\circ$ (corresponding to $k_x = 0.220 \times 2\pi/a$) we get $Z_1 = 1.414$ and $Z_3 = 0.319$. We
should now check whether $\tilde Z_3 \equiv Z_3/Z_1 = 0.225$ lies within the region determined by
the conditions~\eqref{eq:ar:constraints-on-eta-t-3-set-theory} equivalent to the
constraints~\eqref{eq:ar:constraints-on-n} with $n\tsub{min} = 1$ and $n\tsub{max} = 2.51$ (the
maximum index of a medium in which only a single propagative diffraction order exists). Figure
\ref{fig:ar:lens:allowed-region}, in which the value of~$\tilde Z_3$ cited above is marked with
point~$A$, shows that this is indeed the case. Therefore eqs.\ \eqref{eq:ar:d-2} and
\eqref{eq:ar:s:n-2} can be used to calculate the parameters of the AR coating of the crystal:
refractive index $n_2 = 1.649$ and thickness $d_2 = 0.540a$. The geometry of this structure, called
\ref{ar:lens:coating-model-r0} from now on, is shown in fig.\
\ref{fig:ar:lens:coatings-and-binary-gratings}(a) and its reflectance is plotted in fig.\
\ref{fig:ar:lens:coatings-and-binary-gratings}(e) with a solid black line. It can be seen that the
application of the coating reduces significantly the reflectance of the crystal, especially for
small angles of incidence. However, the parameters of~\ref{ar:lens:coating-model-r0} are certainly
not optimal, since its reflectance at the ``design angle'' $\theta = 45^\circ$ is as large as 9\%.
This is due to the relatively large error introduced by the single-mode approximation for
negative-refraction PC bands, as pointed out in ref.~{\onlinecite{SmigajPRB08}}. 

\begin{figure}
  \centering%\floatfont
   \includegraphics{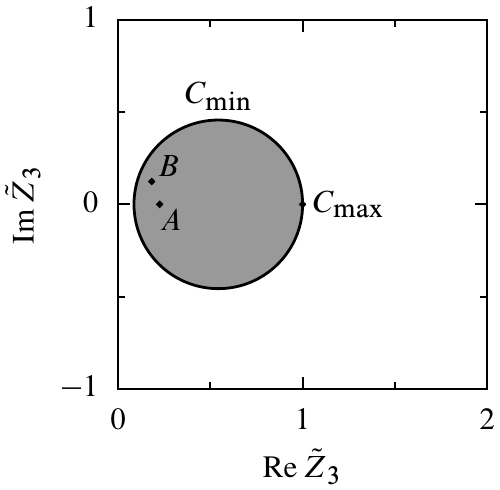}
 
\caption{Shaded circle: region of the complex $\tilde Z_3$ plane determined by the condition
\eqref{eq:ar:constraints-on-eta-t-3-set-theory} equivalent to the constraint
\eqref{eq:ar:constraints-on-n} for $s$~polarization, $n_1 = n\tsub{min} = 1$ and $n\tsub{max} = 2.51$.
Points $A$ and~$B$: reduced impedances $\tilde Z_3$ of structure~\ref{ar:lens:t-14} calculated in
two different ways described in the text.}
  \label{fig:ar:lens:allowed-region}
\end{figure}

We shall evaluate, therefore, an alternative method of obtaining~$Z_3$, which consists in
calculating it directly from the rigorous specular reflection coefficient $r_0$ of the uncoated
crystal at the chosen $\omega$ and~$k_x$. In other words, we assume that $r_0$ can be expressed in
the form $r_0 = (Z_3 - Z_1) / (Z_3 + Z_1)$ [cf.\ eq.~\eqref{eq:ar:fresnel-formulas}] and invert this
formula to obtain $Z_3 = Z_1 (1 + r_0) / (1 - r_0)$. Of course, the effective impedance defined in
this way depends on the material properties of medium~1. Nevertheless, at least for the PC in
question, this dependence is weak for sufficiently small $n_1$: we obtain $Z_3 = 0.258+0.175 i$ for
$n_1 = 1$ and the effective impedance does not change by more than 10\% up to $n_1 = 2.25$. As shown
in fig.\ \ref{fig:ar:lens:allowed-region}, the reduced impedance $\tilde Z_3 = 0.182 + 0.124 i$
corresponding to the above value of $Z_3$ (marked with point~$B$) also lies within the allowed
region of the $\tilde Z_3$ plane. Taking this value of~$Z_3$, from eqs.\ \eqref{eq:ar:d-2} and
\eqref{eq:ar:s:n-2} we get $n_2 = 1.884$ and $d_2 = 0.565a$. The angular dependence of the reflectance
of the PC covered with this coating, shown in fig.\
\ref{fig:ar:lens:coatings-and-binary-gratings}(b) and called \ref{ar:lens:coating-true-r0} in the
following, is plotted in fig.\ \ref{fig:ar:lens:coatings-and-binary-gratings}(e) with a solid gray
line. It is evident that this structure has much better angular tolerance
than~\ref{ar:lens:coating-model-r0}; moreover, its reflectance at $\theta = 45^\circ$ is only
0.05\%. Therefore we choose~\ref{ar:lens:coating-true-r0} as a basis for the further steps of the
algorithm.

\begin{figure}
  \centering
  \includegraphics{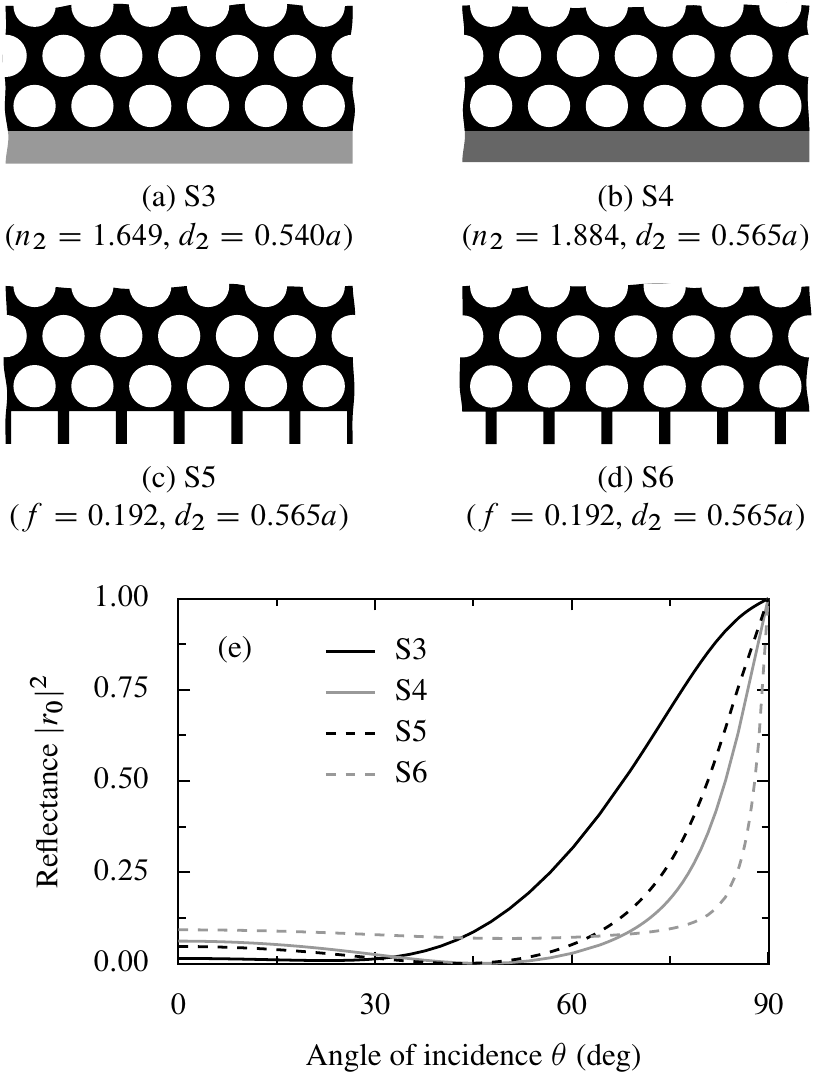}

  \caption{(a)--(b)~Geometry of AR coatings \ref{ar:lens:coating-model-r0} and \ref{ar:lens:coating-true-r0}, characterised by refractive index~$n_2$ and thickness~$d_2$ specified next to the drawings. (c)--(d) Geometry of binary lamellar AR gratings \ref{ar:lens:binary-good} and \ref{ar:lens:binary-bad}, characterised by fill factor~$f$ and thickness~$d_2$ specified next to the drawings. (e)~Angular dependence of the reflectance of the structures shown in parts (a)--(d).}
  \label{fig:ar:lens:coatings-and-binary-gratings}
\end{figure}

%\paragraph{Step 2} 
Numerical inversion of eq.~\eqref{eq:ar:s:eff-eps-2} gives the fill factor $f =
0.192$ of the binary grating mimicking a medium with $n = 1.884$. Since we would like the angular
dependence of $r_0$ to be symmetric with respect to $\theta = 0$, the grating should be positioned
so as to preserve the vertical mirror symmetry axes of the underlying PC. This can be done in two ways, shown in figs.\ \ref{fig:ar:lens:coatings-and-binary-gratings}(c) and~(d). The
reflectance of these two structures, called \ref{ar:lens:binary-good} and \ref{ar:lens:binary-bad},
is plotted in fig.\ \ref{fig:ar:lens:coatings-and-binary-gratings}(e). Clearly,
grating~\ref{ar:lens:binary-good} reproduces fairly faithfully the original reflectance curve of the
AR coating~\ref{ar:lens:coating-true-r0}. On the other hand, grating~\ref{ar:lens:binary-bad}
behaves better in the high-$\theta$ region.

% \paragraph{Step 3} 
The %\fxnote{Change the subscript order to $x1$ etc.}
lamellar gratings obtained in step~2 provide already a remarkable improvement over the uncoated PC
and, in contrast to the AR coatings from step~1, should be manufacturable. Nevertheless, their
geometry can be further ameliorated. To this end, as mentioned in the last paragraph of
section~\ref{sec:ar:algorithm}, we use the Nelder-Mead simplex algorithm to find the optimum values
of the dimensions $w\tsub{i}$, $w\tsub{o}$, $h\tsub{i}$, and $h\tsub{o}$ parametrising the trapezoidal
grating shown in fig.\ \ref{fig:ar:trapezoidal-grating}. The objective function~$\rho$ is defined as
the average of the numerically calculated reflectance of the given structure over the desired
angular tolerance interval $[\theta\tsub{min}, \theta\tsub{max}]$,
\begin{equation}
  \label{eq:ar:objective-function}
  \rho = \frac1{\theta\tsub{max} - \theta\tsub{min}} 
   \int_{\theta\tsub{min}}^{\theta\tsub{max}}
  |{r_0(\theta)}|^2\,d \theta.
\end{equation}
The integral in eq.~\eqref{eq:ar:objective-function} is calculated with the 20-point Gauss-Legendre
quadrature algorithm (ref.\ \onlinecite{NumericalRecipesInC}, section 4.5), whose typical relative accuracy,
${\sim}10^{-5}$, is better than that of the reflectance calculations, ${\sim}10^{-3}$. The initial
shape of the grating is taken to correspond to one of the lamellar gratings obtained in step~2,
i.e., $w\tsub{i} = w\tsub{o} = fa$, $h\tsub{i} = 0$, and $h\tsub{o} = d_2$. The search routine is
terminated when the size of the simplex, defined as the average distance of its vertices from its
geometric centre, falls below~$10^{-5}$. The final values of the geometrical parameters of the
grating are determined by selecting the best among the 16~structures obtained by rounding each of
the parameters delivered by the simplex algorithm upwards or downwards to a multiple of $0.01a$. \begin{figure}
  \centering%\floatfont
   \includegraphics{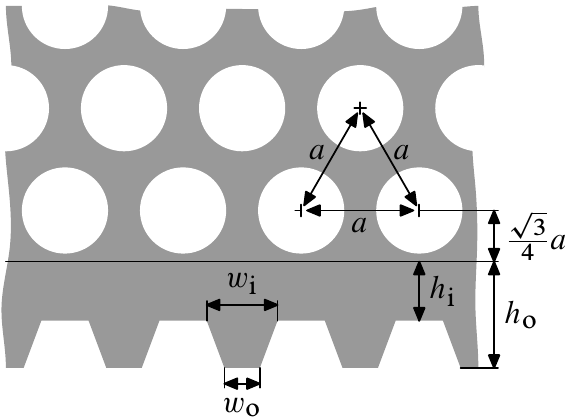}
 
\caption{Definition of the geometrical parameters $w\tsub{i}$, $w\tsub{o}$, $h\tsub{i}$, and $h\tsub{o}$
of a trapezoidal grating superposed on the surface of structure~\ref{ar:lens:t-14}.}
  \label{fig:ar:trapezoidal-grating}
\end{figure}

Application of this procedure with $\theta\tsub{min} = 0^\circ$, $\theta\tsub{max} = 90^\circ$ to
grating~\ref{ar:lens:binary-good} yields grating~\ref{ar:lens:cr112} shown in fig.\
\ref{fig:ar:lens:trapezoidal-gratings}(a). The plot in fig.\
\ref{fig:ar:lens:trapezoidal-gratings}(e) (solid black line) demonstrates the excellent
antireflective properties of this structure (note the scale of the vertical axis). Its average
reflectance is as low as 2.8\%; in fact, $|{r_0(\theta)}|^2$ does not exceed 5.5\% until $\theta =
87^\circ$. The structure does not seem to present special fabrication difficulties---e.g., acute
angles---except possibly for the relatively thin dielectric veins separating the circular holes from
the surface. Should this pose a real experimental difficulty, one can increase the value of $h_
i$ at the expense of a slight performance deterioration. For example, grating \ref{ar:lens:cr115}
with $h\tsub{i} = 0.08a$ {[fig.\ \ref{fig:ar:lens:trapezoidal-gratings}(b)]} has average reflectance of 4.8\%. 
\begin{figure}
  \centering
  \includegraphics{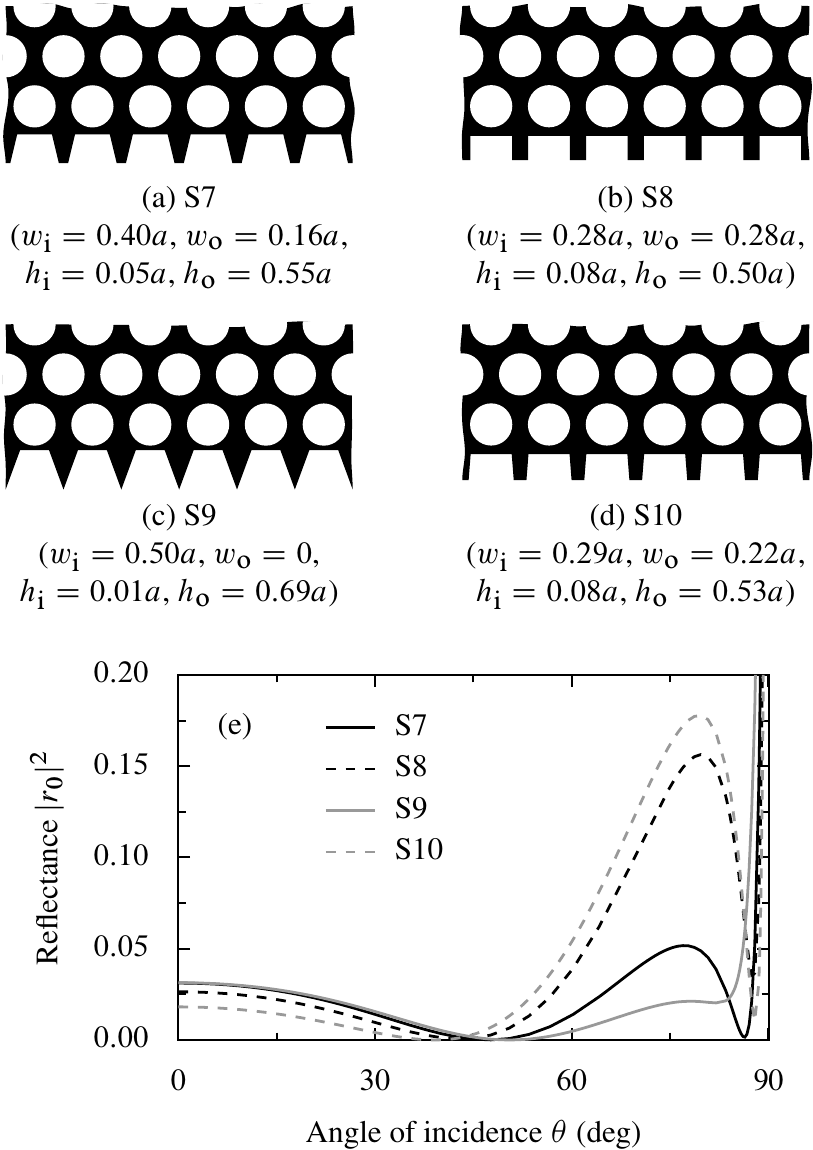}

\caption{{(a)--(d)}~Geometry of AR gratings \ref{ar:lens:cr112}--\ref{ar:lens:cr111} characterized
by parameters $w\tsub{i}$, $w\tsub{o}$, $h\tsub{i}$, and $h\tsub{o}$ specified next to the drawings.
{(e)}~Angular dependence of the reflectance of the structures shown in parts (a)--(d). To
help visualize the details of the $|{r_0(\theta)}|^2$ dependence, the $y$~axis has been truncated
at $|{r_0}|^2 = 0.2$.}
  \label{fig:ar:lens:trapezoidal-gratings}
\end{figure}
Figure \ref{fig:ar:lens:field-maps-gratings} shows the map of the modulus of the electric field
produced by a point source placed above a PC slab coated with AR gratings of type
\ref{ar:lens:cr112} from above and below.
The comparison with fig.\ \ref{fig:ar:lens:field-maps-uncoated} reveals the significant improvement
brought about by the AR grating: not only are the reflected beams prominent in the upper part of the
latter figure suppressed, but the amplitude of the image formed by the lens grows to 159\,V$/$m,
which is two times better than in the situation from fig.\ \ref{fig:ar:lens:field-maps-uncoated}(b).\begin{figure}
  \centering%\floatfont
   \includegraphics{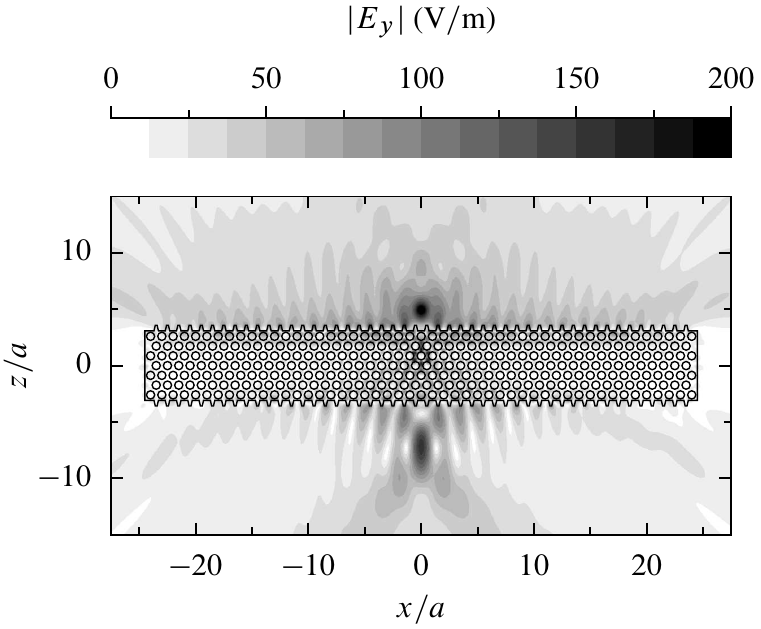}
 
\caption{Modulus of the electric field generated by an $s$-polarized wire source with current 1\,A
located above a slab of the PC studied in section~\ref{sec:ar:example-flat-lens} with
\ref{ar:lens:cr112}-type gratings placed on its horizontal surfaces.}
  \label{fig:ar:lens:field-maps-gratings}
\end{figure}

Optimization of structure~\ref{ar:lens:binary-bad} leads to gratings with average reflectance
comparable to that of \ref{ar:lens:cr112} and \ref{ar:lens:cr115} but composed of ``narrower''
trapezoids ($w\tsub{i} + w\tsub{o} \approx 0.2a$), thus less suitable for fabrication. Therefore we
omit the detailed discussion of these structures.

As a complement, we mention that in refs.\ \onlinecite{SmigajSPP4} and \onlinecite{SmigajOL09} two
other trapezoidal AR gratings, here denoted \ref{ar:lens:cr108} and \ref{ar:lens:cr111}, were
presented. Their geometrical parameters, shown in figs.\ \ref{fig:ar:lens:trapezoidal-gratings}(c)
and~(d), were obtained by minimizing the objective function $\frac{2}{\pi} \int_0^{\pi/2}
|{r_0(\theta)}|\,d \theta$ (average modulus of the specular reflection coefficient~$r_0$) calculated
with a less accurate quadrature algorithm. The average reflectance of structure \ref{ar:lens:cr108}
in the full $0^\circ$--$90^\circ$ range, 2.8\%, matches that of \ref{ar:lens:cr112}; in a more
restricted range, say, $0^\circ$--$80^\circ$, the performance of grating \ref{ar:lens:cr108} is even
slightly better. Nonetheless, its disadvantage lies in the presence of very thin dielectric veins at
the surface. Grating \ref{ar:lens:cr111}, with $h\tsub{i} = 0.08a$, is devoid of this problem. However,
it is superseded by structure \ref{ar:lens:cr115} with identical~$h\tsub{i}$, which has somewhat lower
average reflectance.

Finally, a word about tolerance to fabrication imperfections. A fabrication process invariably
perturbs the geometrical parameters of the manufactured structure. 
% Therefore it is important to verify the sensitivity of the proposed gratings to such perturbations. 
To assess the sensitivity of
the proposed gratings to fabrication errors, we have determined the maximum perturbation of each of
the four geometrical parameters of grating \ref{ar:lens:cr112} for which the grating's average
reflectance in the angular range $0 \leq \theta \leq \theta\tsub{max}$ did not exceed 5\%. Three
values of $\theta\tsub{max}$ were considered: $90^\circ$, $80^\circ$, and $60^\circ$. The results of
this test are summarized in table \ref{tab:ar:lens:tolerance-cr112}. It can be seen that the grating
is more sensitive to variations of the height of the trapezoids (via the $h\tsub{o} $ and $h\tsub{i}$
parameters) than of their width ($w\tsub{o} $ and $w\tsub{i}$). The constraints for $\theta\tsub{max} =
90^\circ$ and $\theta\tsub{max} = 80^\circ$ are rather stringent and unlikely to be met in practice.
In contrast, fabrication of a structure satisfying the constraints for $\theta\tsub{{max}} = 60^\circ$
seems well within reach of current technology. 
We have also tested the frequency tolerance of grating \ref{ar:lens:cr112}, finding that the its
average reflectance stays below 5\% for $0.3094 \leq \omega a/2\pi c \leq 0.3113$ ($\theta\tsub{max}
= 90^\circ$), $0.3048 \leq \omega a/2\pi c \leq 0.3122$ ($\theta\tsub{max} = 80^\circ$), and $0.2511
\leq \omega a/2\pi c \leq 0.3173$ ($\theta\tsub{max} = 60^\circ$). This tolerance seems quite
sufficient for applications related to lensing.

\begin{table*}
  \centering%\floatfont
  \begin{tabular}{l@{\hspace{1em}}c@{\hspace{1em}}c@{\hspace{1em}}c@{\hspace{1em}}c}
    \hline
    $\theta\tsub{max}$ & 
    $w\tsub{i}$ & $w\tsub{o}$ & $h\tsub{i}$ & $h\tsub{o}$ \\ 
    \hline
    $90^\circ$&
    $0.378$--$0.424a$ (\hphantom{0}22\,nm)&
    $0.151$--$0.170a$ (\hphantom{0}9\,nm)&
    $0.045$--$0.055a$ (\hphantom{0}5\,nm)&
    $0.543$--$0.557a$ (\hphantom{0}7\,nm)\\
    $80^\circ$&
    $0.330$--$0.463a$ (\hphantom{0}63\,nm)&
    $0.132$--$0.186a$ (26\,nm)&
    $0.033$--$0.062a$ (14\,nm)&
    $0.530$--$0.569a$ (19\,nm)\\
    $60^\circ$&
    $0.192$--$0.533a$ (162\,nm)&
    $0.084$--$0.208a$ (59\,nm)&
    $0\hphantom{.000}$--$0.082a$ (39\,nm)&
    $0.497$--$0.586a$ (42\,nm)\\
    \hline
  \end{tabular}
\caption{Ranges of geometrical parameters of grating \ref{ar:lens:cr112} for which its average
reflectance at frequency $0.311\times 2\pi c/a$ in the angular range $0 \leq \theta \leq
\theta\tsub{max}$ does not exceed 5\%. The numbers in parentheses are the lengths of the tolerance
intervals for $a = 476$\,nm, which corresponds to operation wavelength $\lambda = a/0.311 =
1530$\,nm. Note that the tolerance intervals correspond to perturbations of \emph{one parameter at a
time} (\emph{not} all parameters simultaneously).}
  \label{tab:ar:lens:tolerance-cr112}
\end{table*}

\section{Application to a supercollimating photonic crystal}
\label{sec:ar:example-self-collimation}

The second example to be considered is a PC composed of a square lattice of air holes of radius $r =
0.3a$, where $a$~is the lattice constant, etched in a dielectric matrix of permittivity $\epsilon =
12.25$. Near the frequency $\omega = 0.265 \times 2\pi c/a$ its EFCs for $p$~polarization take a
square-like shape (cf.\ fig.\ \ref{fig:ar:coll:efc}). In consequence, supercollimated
beams \cite{KosakaPRB98,PratherJPD07} can propagate in the crystal.

\begin{figure}
  \centering%\floatfont
  \includegraphics{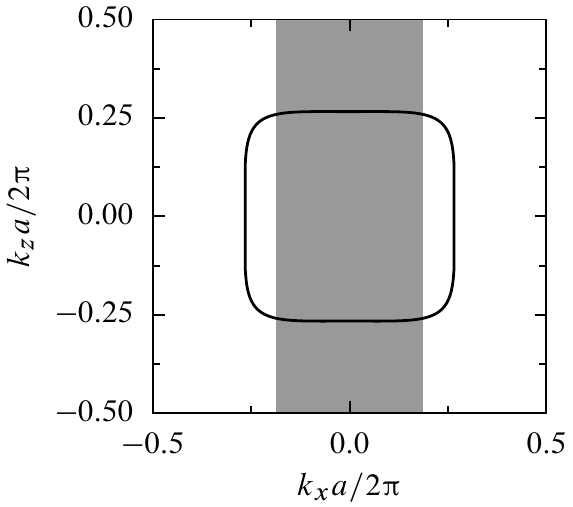}
\caption{EFC of the PC studied in section~\ref{sec:ar:example-self-collimation} at frequency $\omega
= 0.265 \times 2\pi c/a$. The shaded region corresponds to the range $\theta \leq 45^\circ$
($|{k_x}| \leq 0.187 \times 2\pi/a$), where the EFC is approximately flat and for which the
minimization of the PC's reflectance is made.}
  \label{fig:ar:coll:efc}
\end{figure}

{The solid and dashed black curves in fig.~\ref{fig:ar:coll:bare-coatings}(e) show the angular dependence of the reflectance of this PC at
$\omega = 0.265 \times 2\pi c/a$ and for two different truncation planes, corresponding to the structures shown in figs.\
\ref{fig:ar:coll:bare-coatings}(a)--(b) and called \ref{ar:coll:uncut-holes} and \ref{ar:coll:cut-holes}
in the following.} It is seen that the crystal cut through hole centers has fairly
low reflectance: about 10\% at normal incidence and decreasing for larger angles up to $\theta
\approx 65^\circ$. This level of power losses might in fact be already sufficient for practical
applications. Nevertheless, for the sake of illustration, we shall present the design procedure of
AR gratings that help to decrease even further the reflectance of the PC in question.

\begin{figure}
  \centering
  \includegraphics{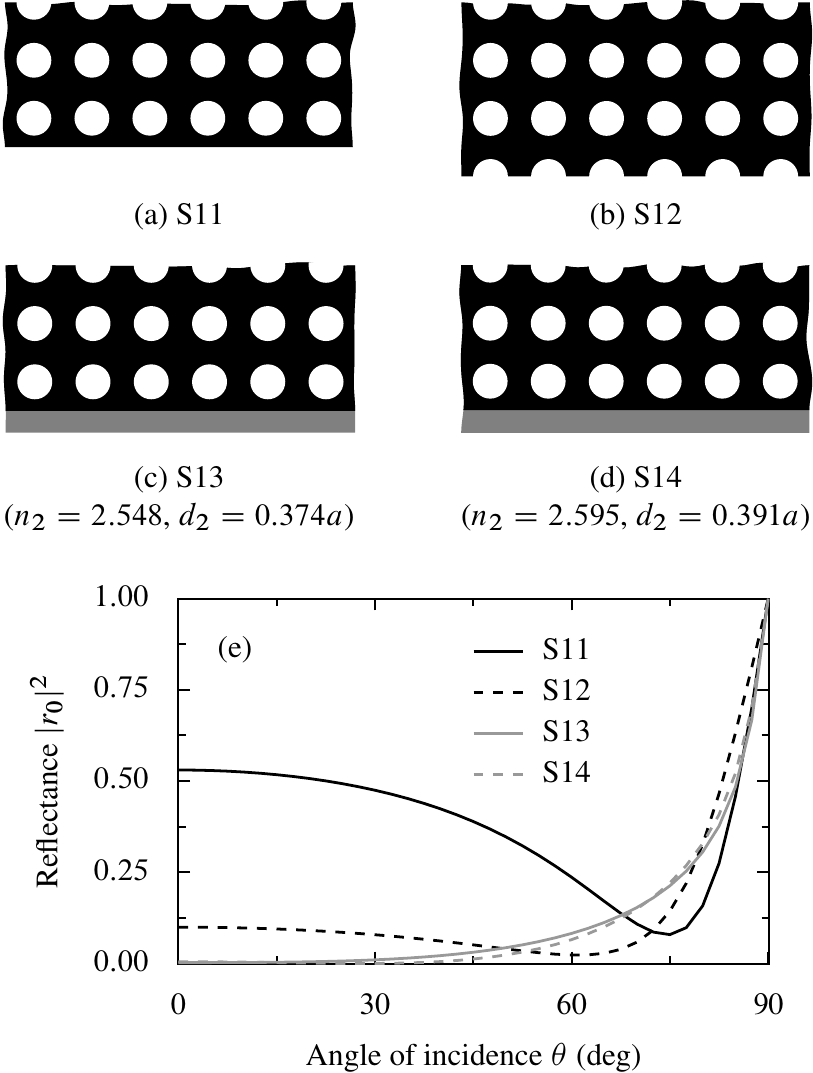}

  \caption{{(a)--(b)} Geometry of the PC studied in section~\ref{sec:ar:example-self-collimation} truncated along
a plane (a) lying midway between two neighboring rows of holes, (b)
crossing the centers of holes.
{(c)--(d)}~Geometry of AR coatings \ref{ar:coll:coating-model-r0} and
    \ref{ar:coll:coating-true-r0}, characterized by refractive index~$n_2$ and thickness~$d_2$
    specified next to the drawings. {(e)}~Angular dependence of the reflectance of the structures
    shown in parts {(a)--(d)}.}
  \label{fig:ar:coll:bare-coatings}
\end{figure}

% \paragraph{Step 1} 

Here, we are mostly interested in coupling the incoming light to modes lying on the flat horizontal
part of the PC's EFC. As shown in fig.\ \ref{fig:ar:coll:efc}, at frequency $\omega = 0.265 \times
2\pi c/a$ this corresponds roughly to the range $|\theta| \leq 45^\circ$, i.e., $|{k_x}| \leq 0.187
\times 2\pi/a$. Therefore we choose $\theta = 22.5^\circ$ as the design angle of the AR coating. As
in the previous section, we test two different ways of calculating the effective immittance (in this
case, admittance) of the crystal cut along a constant-permittivity plane, i.e., structure
\ref{ar:coll:uncut-holes}. The effective-medium model presented in {ref.\ \onlinecite{SmigajPRB08}} yields
$Y_3 = 6.138$. In turn, the effective admittance calculated from the rigorous reflection coefficient
of structure \ref{ar:coll:uncut-holes} embedded in air is $Y_3 = 6.075-1.191 i$. Figure\
\ref{fig:ar:coll:allowed-region} shows that the reduced admittances corresponding to both these
values lie within the region of the complex $\tilde Y_3$ plane determined by the
conditions~\eqref{eq:ar:constraints-on-eta-t-3-set-theory} equivalent to the
constraints~\eqref{eq:ar:constraints-on-n} with $n_1 = n\tsub{min} = 1$, $n\tsub{max} = 3.391$ (the
refractive index for which the second propagative diffraction order appears) and $\theta =
22.5^\circ$. The parameters of the AR coatings determined from these two values of~$Y_3$ are ($n_2 =
2.548$, $d_2 = 0.374$) and ($n_2 = 2.595$, $d_2 = 0.391$), respectively. {Figures
\ref{fig:ar:coll:bare-coatings}(c)--(d) show} the geometry of these coatings, henceforth referred to as
\ref{ar:coll:coating-model-r0} and \ref{ar:coll:coating-true-r0}, {whereas the angular dependence of
their reflectance is plotted with solid and dashed gray lines in fig.\ \ref{fig:ar:coll:bare-coatings}(e)}. As in the PC lens case, the AR coating \ref{ar:coll:coating-true-r0} designed
using the value of $Y_3$ obtained from the rigorous reflection coefficient of the crystal performs
slightly better than the other one. Therefore structure \ref{ar:coll:coating-true-r0} shall be used
in the subsequent design step.

\begin{figure}
  \centering%\floatfont
  \includegraphics{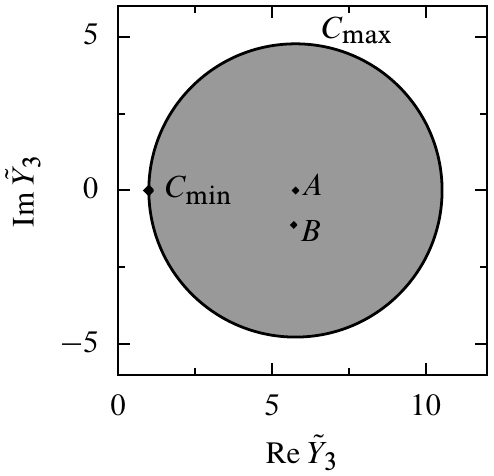}
 
\caption{Shaded circle: region of the complex $\tilde Y_3$ plane determined by the condition
\eqref{eq:ar:constraints-on-eta-t-3-set-theory} equivalent to the constraint
\eqref{eq:ar:constraints-on-n} for $p$~polarization, $n_1 = n\tsub{min} = 1$ and $n\tsub{max} =
3.391$. Points $A$ and~$B$: reduced admittances $\tilde Y_3$ of structure~\ref{ar:coll:uncut-holes}
calculated in two different ways described in the text.}
  \label{fig:ar:coll:allowed-region}
\end{figure}

% \paragraph{Step 2} 

From numerical inversion of eq.~\eqref{eq:ar:p:eff-eps-2} it follows that the fill factor of the
binary grating mimicking a medium with $n = 2.595$ for $p$~polarization is $f = 0.812$. Figures
\ref{fig:ar:coll:gratings}(a)--(b) show the geometry of the two gratings, called
\ref{ar:coll:grating-normal-pos} and \ref{ar:coll:grating-shifted-pos}, with this fill factor and a
mirror symmetry plane perpendicular to the direction of periodicity. From the juxtaposition of their
reflectance curves [fig.\ \ref{fig:ar:coll:gratings}(c)] it follows that structure
\ref{ar:coll:grating-normal-pos} has somewhat better performance than
\ref{ar:coll:grating-shifted-pos}. {In fact,} there is some similarity between the geometry of grating
\ref{ar:coll:grating-normal-pos} and the truncated crystal \ref{ar:coll:cut-holes}, which also
exhibited fairly low reflectance: {t}he surface of both these structures contains ``teeth'' shifted by
$\frac12a$ in the horizontal direction with respect to the positions of the circular holes.
Therefore, one could view the crystal \ref{ar:coll:cut-holes} as an imperfect realization of the AR
grating \ref{ar:coll:grating-normal-pos}.

\begin{figure}
  \centering
  \includegraphics{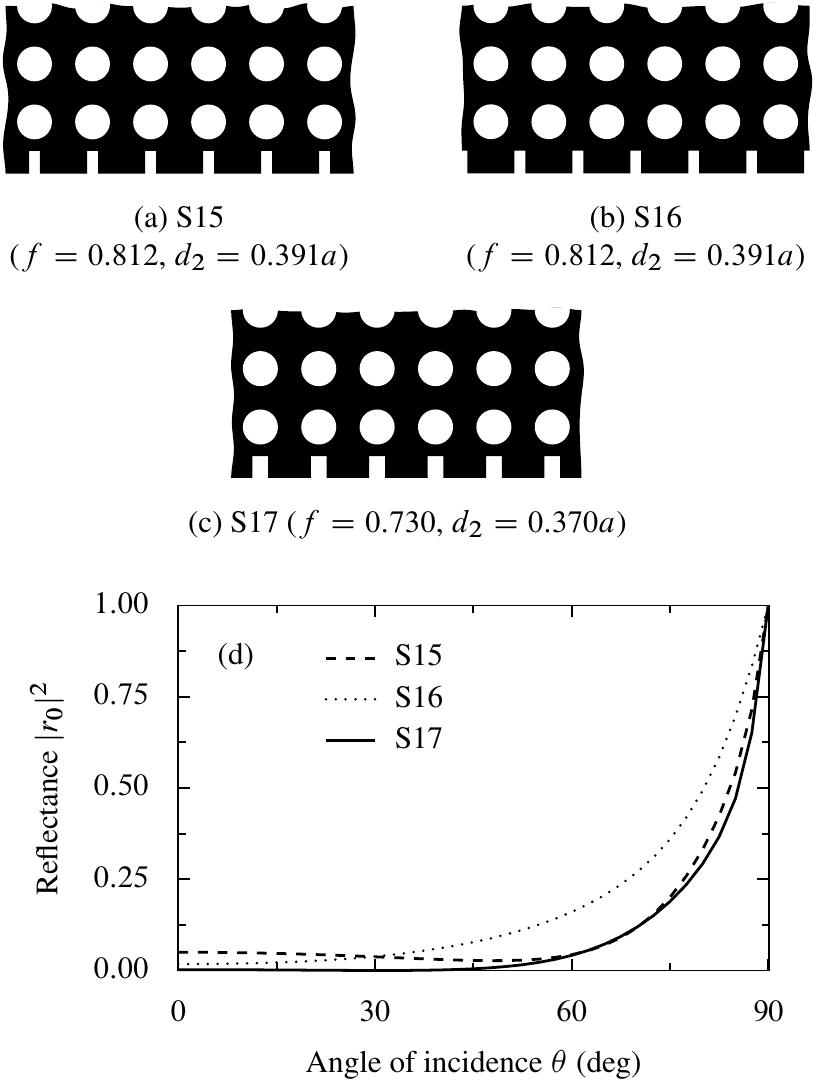}
  
\caption{{(a)--(c)}~Geometry of binary lamellar AR gratings \ref{ar:coll:grating-normal-pos},
\ref{ar:coll:grating-shifted-pos} and \ref{ar:coll:grating-optimised} characterized by fill
factor~$f$ and thickness~$d_2$ specified next to the drawings. {(d)}~Angular dependence of the
reflectance of the structures shown in parts (a)--(c).}
  \label{fig:ar:coll:gratings}
\end{figure}

%\paragraph{Step 3} 

The lamellar grating \ref{ar:coll:grating-normal-pos} can be further ameliorated by adjusting its
thickness~$d_2$ and fill factor~$f$ to minimize the objective function~$\rho$ defined in eq.\
\eqref{eq:ar:objective-function}. We take $\theta\tsub{min} = 0$, $\theta\tsub{max} = 45^\circ$ and,
as before, perform the optimization with the Nelder-Mead simplex algorithm. This leads to structure
\ref{ar:coll:grating-optimised} with $d_2 = 0.37a$ and $f = 0.73$, shown in fig.\
\ref{fig:ar:coll:gratings}(c). Its reflectance curve is plotted in fig.\
\ref{fig:ar:coll:gratings}(d) (solid line). In the angular range $0 \leq \theta \leq 45^\circ$, the
reflectance never exceeds 0.6\%, on average amounting to only 0.12\%. The structure does not seem to
present any special fabrication problems. It is possible to continue the grating's optimization by
allowing it to take a trapezoidal rather than a lamellar shape; however, in view of its already very
good AR properties, this appears unnecessary.

\section{{Application to a} non-reciprocal mirror}
\label{sec:ar:example-superprism}

{In this section, the procedure is applied to} a unidirectional mirror similar to that proposed
in ref.\ \onlinecite{VanwolleghemPRB09}. This device has the form of a slab of the PC shown in {the inset of} fig.\
\ref{fig:ar:shamrock-squares}(a). The PC consists of a hexagonal lattice of non-centrosymmetric
motifs etched in a magneto-optical matrix characterized by a gyrotropic permittivity tensor
\begin{equation}
  \label{eq:ar:gyrotropic-permittivity}
  \epsilon = 
  \begin{bmatrix}
    (2.25)^2 & 0 & 0.1 i \\
    0 & (2.25)^2 & 0 \\
    -0.1 i & 0 & (2.25)^2 
  \end{bmatrix} .
\end{equation}
The presence of this {magneto-optical} material lifts the time-reversal symmetry of Maxwell's equations. Since the
spatial inversion symmetry is also broken, the crystal becomes nonreciprocal: the usual property of
the dispersion relation, $\omega(k) = \omega(-k)$ {(ref.\ \onlinecite{SakodaBook01}, pp.\ 22--23)}, no longer holds.
The $p$-polarization EFC of this crystal at frequency $\omega = 0.4537 \times 2\pi c/a$ is shown in
fig.\ \ref{fig:ar:shamrock-squares}({a}). It can be seen that plane waves impinging at the angle of
about $44^\circ$ (corresponding to $k_x = 0.315\times 2\pi/a$) on the bottom surface of a slab made
from this PC will be coupled to its propagative mode, and thus will be partially transmitted.
However, waves travelling in the opposite direction, incident from the top, will be totally
reflected, since the crystal has no propagative modes with $-0.323\times 2\pi/a \leq k_x \leq
-0.308\times 2\pi/a$. This behaviour justifies the name \emph{unidirectional mirror}.
The problem with the presented device is its large forward loss: even waves propagating in the
``allowed'' direction undergo a significant reflection on the surfaces of the slab. This can be seen
in fig.\ \ref{fig:ar:shamrock-squares}({b}), where the reflectance curve of the crystal truncated as
in fig.\ \ref{fig:ar:shamrock-squares}(a) is juxtaposed with the relevant fragment of the EFC.
Clearly, to be useful in practice, the unidirectional mirror needs to be coated with some AR
structure. 

\begin{figure}
  \centering
  \includegraphics{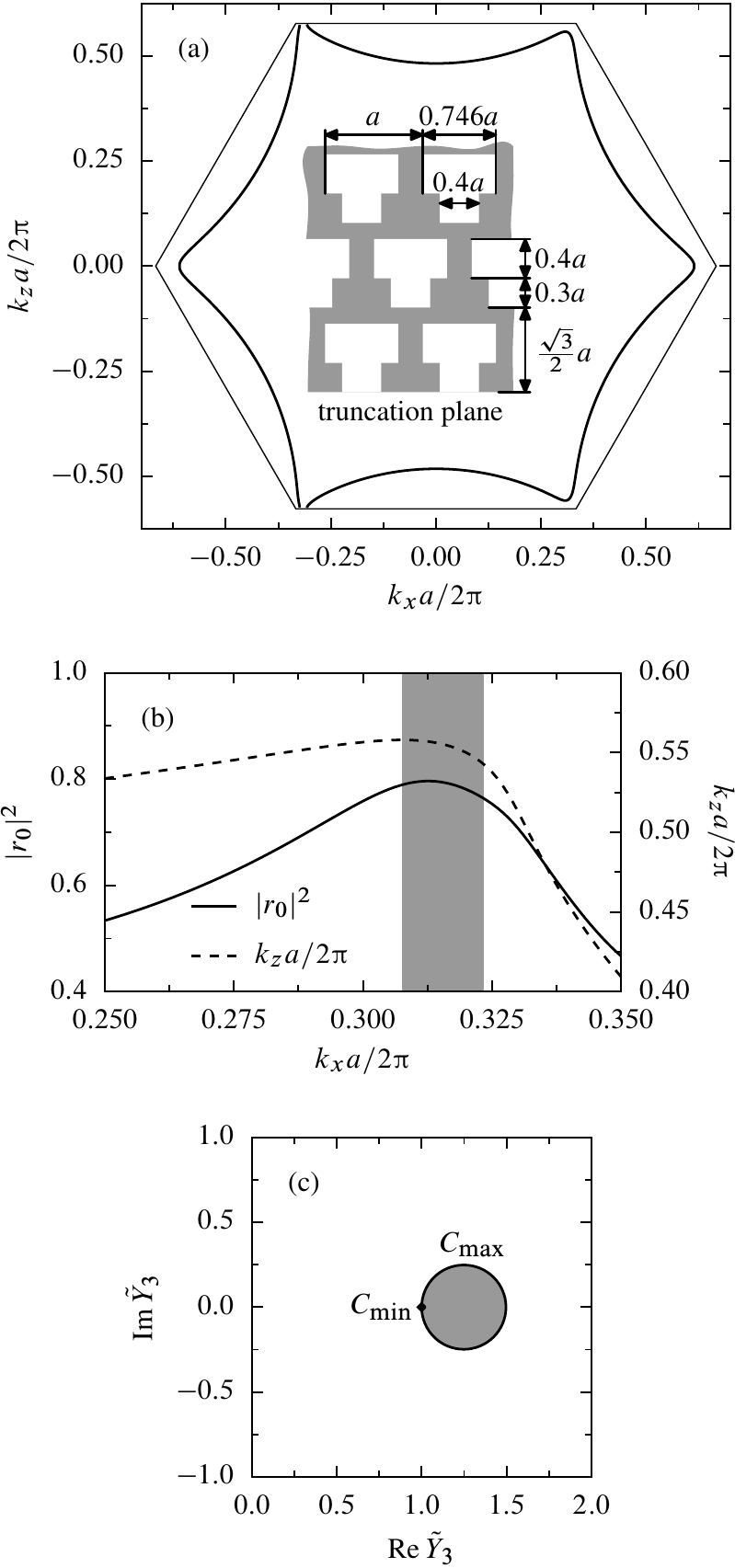}

\caption{(a)~$p$-polarization EFC of the magneto-optical PC shown in the inset at frequency $\omega = 0.4548
\times 2 \pi c/a$. (b)~Solid line: $k_x$-dependence of the reflectance $|r_0|^2$ of this crystal,
placed in air and truncated in the way indicated in the inset of part~(a). Dashed line: a fragment of the EFC
from part~(a). The shaded region indicates the range of $k_x$ for which the crystal has a
non-reciprocal gap, i.e.\ there are no propagative modes with the $x$ component of the Bloch vector
equal to $-k_x$. (c)~Shaded circle: region of the complex $\tilde Y_3$ plane determined by the
condition \eqref{eq:ar:constraints-on-eta-t-3-set-theory} equivalent to the constraint
\eqref{eq:ar:constraints-on-n} for $p$~polarization, $n_1 = n\tsub{min} = 1$ and $n\tsub{max} =
1.50$. The reduced admittance of the crystal, $\tilde Y_3 = 13.5-7.3i$, lies far beyond the range of
the graph.}
  \label{fig:ar:shamrock-squares}
\end{figure}

Unfortunately, the design of an appropriate AR grating using the procedure described in
section~\ref{sec:ar:algorithm} turns out to be impossible. For instance, at $k_x = 0.315\times 2\pi
c/a$, near the centre of the region where the unidirectional mirroring effect occurs, the reflection
coefficient of the uncoated PC is $r_0 = 0.900-0.055i$, which corresponds to $\tilde Y_3 =
13.5-7.3i$. As evidenced by fig.\ \ref{fig:ar:shamrock-squares}(d), this lies far outside the region
of the complex $\tilde Y_3$ plane determined by the
conditions~\eqref{eq:ar:constraints-on-eta-t-3-set-theory} equivalent to the
constraints~\eqref{eq:ar:constraints-on-n} with $n_1 = n\tsub{min} = 1$ and $n\tsub{max} = 1.50$
(the refractive index of the most optically dense coating in which the second propagative
diffraction order would not yet appear). In fact, if we blindly apply eqs.\
\eqref{eq:ar:necessary-cond-simplified} and \eqref{eq:ar:p:n-2} to calculate the refractive index of
the optimum AR coating, we obtain $n_2 = 5.79$ or $0.70$. It is obvious that none of these indices
can be simulated by any binary grating composed of the constituent materials of the PC. Therefore,
an AR structure for the unidirectional mirror will probably need to be designed with some purely
numerical method. In particular, {Lawrence \textit{et al.}} \cite{LawrencePRA09} have shown their approach to give good results
for a superprism that, uncoated, has an extremely high reflectance ($|r_0|^2 = 0.996$).

\section{Evanescent waves regime}
\label{sec:evanescent}

The behaviour of evanescent waves is approached through the example of the photonic crystal flat
lens described in section \ref{sec:ar:example-flat-lens}. In that case, the amplification of these
waves is of vital importance to obtain subwavelength resolution \cite{PendryPRL00}. In a photonic
crystal flat lens, two mechanisms may lead to an amplification of evanescent waves \cite{LuoPRB03}:
the \textit{single interface resonances} and the \textit{overall resonances} (see ref.\
{\onlinecite{LuoPRB03}} for discussion in detail). While the first mechanism is actually involved in the
original flat lens, the solely second mechanism has been investigated. This situation can be changed
thanks to the new models introduced recently \cite{SmigajPRB08,LawrencePRA09,SmigajAPL11} and based
on reflectivity on semi-infinite crystals.

The present approach is based on the following conjecture: \textit{assuming that a nearly
constant effective index is obtained over all the range of wavevectors corresponding to propagative
waves, it can be expected that this remains true when this range is continued in the evanescent
regime}. For instance, this behaviour has already been observed in multilayers \cite{PierreJMO07}.
For the photonic crystal flat lens, the refractive index is close to $-1$ at $\omega = 0.311 \times 2 \pi c
/ a$ (from the dispersion law of Fig. \ref{fig:ar:lens:efc}) and, with the {AR} grating, it can
be considered that the effective index is nearly $-1$ for the range of propagating waves. Now, if
the effective index remains nearly $-1$ for evanescent waves, then it can be expected that
\textit{single\wojtek{-}interface resonances} are present around the working frequency $\omega = 0.311 \times 2 \pi
c / a$.

\begin{figure}                                               
  \centering %\floatfont                                                      
  \includegraphics{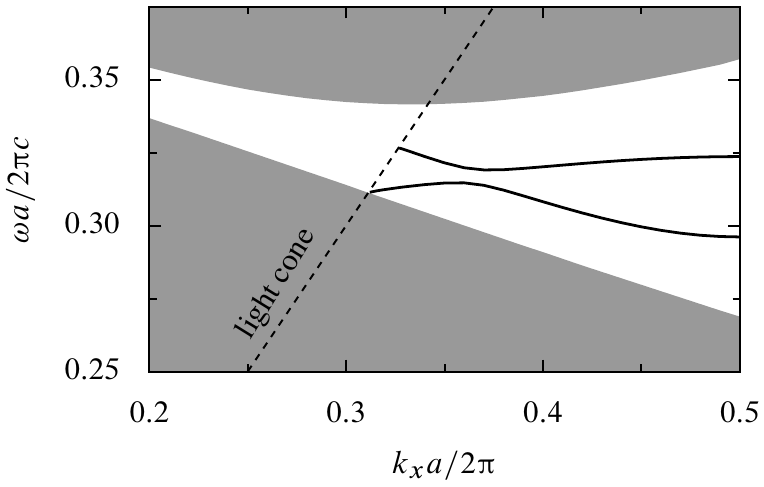}
\caption{Single\wojtek{-}interface resonances of the structure \ref{ar:lens:cr112}. The shaded areas are the bulk bands of the underlying PC.}
  \label{fig:ar:s-modes-S7}
\end{figure}

Fig. \ref{fig:ar:s-modes-S7} shows the single{-}interface resonances of structure \ref{ar:lens:cr112} below the light cone. It
is observed that the {dispersion} curve of these resonances {starts} from the intersection between the {bands of the bulk PC} and the the light cone at $\omega = c k_x = 0.311 \times 2 \pi c / a$. This shows that the
effective index has no jump from propagative to evanescent waves, and thus confirms the conjecture
proposed above, as well as the result found in {ref.\ \onlinecite{PierreJMO07}} in the one-dimensional case. To our
knowledge, no model can predict this continuity of the effective index at this change of regime. This might be a subject for furter investigations.

% It has to be noticed that the curve of the resonances cannot be considered as the horizontal line
% corresponding to $\omega = 0.311$ $2 \pi c / a$. However the obtained structure with optimized
% antireflection grating may be used as an initial condition in a second optimization procedure in
% order to obtain a curve of resonances as flat as possible. Fig. \ref{fig:ar:s-modes-S7} shows the
% dispersion curve of resonances after a second optimization with criterion of flatness for the range
% of wavevectors from $k_x = 0.45 (2 \pi / a)$ to $k_x = 0.5 (2 \pi / a)$. The resulting curve
% indicates that the procedure presented here may be used to define new guiding structure
% presenting low group velocity and slow light propagation \cite{Philippe}. 

% \begin{figure*}                                               
%   \centering %\floatfont                                                      
%   \includegraphics{pic/cr201-surface-modes-N-5-zoom}
% \caption{\textit{Single interface resonances} (full circles) of the optimized structure and Bloch modes
% bands (hatched domain) of the original crystal S1.}
%   \label{fig:ar:s-modes-SO}
% \end{figure*}

\section{Conclusion}
\label{sec:conclusion}

In this article we have presented a new method of designing gratings that, superimposed on surfaces
of PCs crystals, will minimize their reflectance. The design algorithm consists of three steps.
First, the parameters of a homogeneous-layer AR coating are calculated from an effective-medium
approximation of the PC in question. Second, an analytical effective-medium theory of gratings is
used to find the parameters of a binary lamellar grating composed solely of the constituent
materials of the crystal and approximating the coating obtained in the previous step. Third, the
shape of the grating is refined with a numerical local-search routine so as to minimize the
reflectance of the structure in the desired angular or frequency range. This last step is necessary
owing to the approximations made in the analytical derivations used in the first two steps of the
procedure.

This algorithm of AR grating design can be viewed as complementary to the method proposed by
Lawrence \textit{et al.} \cite{LawrenceAPL08,LawrencePRA09}. While their approach is based on an exhaustive scan of the whole
parameter space (made very efficient by the application of the matrix-valued effective immittance of
gratings), ours rests on approximate analytical considerations used to find a starting point for a
\emph{local} search procedure.

The proposed method has been applied to three example crystals with
EFCs of different curvature: a supercollimating crystal with a very flat EFC, a crystal exhibiting
negative refraction, with almost circular EFC, and a PC superprism, whose EFC has a kink. In the two
first cases, the design process succeeded in producing AR gratings ensuring very low reflectance in
a wide angular range. The obtained structures are quite compact and apparently rather
straightforward to fabricate. In the last case, the procedure broke down owing to the violation of
the constraints~\eqref{eq:ar:constraints-on-eta-t-3-set-theory} on the effective immittance of the
crystal that must be satisfied in order that the AR coating produced in the first step can be
approximated with a binary grating made of realistic materials. The existence of these constraints
is the basic limitation of the presented procedure. 

\appendix
\section{Constraints on~$\Xi_3$}
\label{appendixA}
We begin by noting that the conditions~\eqref{eq:ar:constraints-on-n} can always be rewritten in the
form
\begin{equation}
  \label{eq:ar:constraints-on-eta-t-2}
  \Xi\tsub{min}^2 \leq \Xi_2^2 \leq \Xi\tsub{max}^2
\end{equation}
with appropriate $\Xi\tsub{min}$ and $\Xi\tsub{max}$. Specifically, for $s$ polarization, the formulas for
$Z\tsub{min}^2$ and~$Z\tsub{max}^2$ follow readily from eq.~\eqref{eq:ar:s:n-2}:
\begin{equation}
  \label{eq:ar:s:eta-t-2-bounds}
\begin{array}{l}
  Z\tsub{min}^2 = \displaystyle\frac{1}{n\tsub{max}^2-n_1^2\sin^2\theta} \, ,
  \\[4mm]
  Z\tsub{max}^2 = \displaystyle\frac{1}{n\tsub{min}^2-n_1^2\sin^2\theta} \,.
\end{array}
\end{equation}
For $p$ polarization, due to the more complex form of eq.~\eqref{eq:ar:p:n-2} and the presence of
the supplementary condition \eqref{eq:ar:p:constraint-on-Y-2}, several cases must be considered. The
final formulas for $Y\tsub{min}^2$ and~$Y\tsub{max}^2$ are given in
table~\ref{tab:ar:p:eta-t-2-bounds}.

\begin{table*}
  \centering%\floatfont
  \newcommand{\nmin}{n\tsub{min}}
  \newcommand{\nmax}{n\tsub{max}}
  \begin{tabular}{lll}
    \hline
    Range of $\nmin^2$, $\nmax^2$ and $n_x^2$ & 
    $Y\tsub{min}^2$ & 
    $Y\tsub{max}^2$ \\
    \hline    
    $0<n_x^2\leq \frac{\nmin^2}{2} \, ,\hspace{10mm} 
    \nmin^2 \leq \nmax^2$ &
    $\frac{\nmin^4}{\nmin^2-n_x^2}$ \hspace{10mm} & 
    $\frac{\nmax^4}{\nmax^2-n_x^2}$ \\[2\jot]
    $\frac{\nmin^2}{2}<n_x^2<\nmin^2\leq \nmax^2\leq 2 n_x^2$ & 
    $\frac{\nmax^4}{\nmax^2-n_x^2}$ \hspace{10mm} & 
    $\frac{\nmin^4}{\nmin^2-n_x^2}$ \\[2\jot]
    $\frac{\nmin^2}{2}<n_x^2<\nmin^2 \, , \hspace{10mm}
    2 n_x^2<\nmax^2\leq \frac{\nmin^2 n_x^2}{\nmin^2-n_x^2}$ \hspace{10mm} &
    $4 n_x^2$ &
    $\frac{\nmin^4}{\nmin^2-n_x^2}$ \\[2\jot]
    $ \frac{\nmin^2}{2}<n_x^2<\nmin^2 \, , \hspace{10mm}
    \frac{\nmin^2 n_x^2}{\nmin^2-n_x^2}<\nmax^2$ \hspace{10mm} & 
    $ 4 n_x^2$ & 
    $\frac{\nmax^4}{\nmax^2-n_x^2}$ \\[2\jot]
    $\nmin^2 \leq n_x^2<\nmax^2\leq 2 n_x^2$ & 
    $\frac{\nmax^4}{\nmax^2-n_x^2}$ \hspace{10mm} & 
    $ \infty$\\[2\jot]
    $\nmin^2 \leq n_x^2 \, , \hspace{10mm} 2 n_x^2 < \nmax^2$ & 
    $4 n_x^2$ & 
    $ \infty$\\
    \hline
  \end{tabular}
\caption{Minimum and maximum bounds on $Y_2^2$ sufficient and necessary for fulfillment of the
condition \eqref{eq:ar:constraints-on-n} together with the constraint
\eqref{eq:ar:p:constraint-on-Y-2} for at least one of the solutions \eqref{eq:ar:p:n-2} of
eq.~\eqref{eq:ar:p:n-2-quadratic}. The symbol $n_x^2$ denotes $n_1^2\sin^2 \theta$.}
  \label{tab:ar:p:eta-t-2-bounds}
\end{table*}

To arrive at the form of the constraints on~$\Xi_3$, we substitute
eq.~\eqref{eq:ar:necessary-cond-simplified} into inequality~\eqref{eq:ar:constraints-on-eta-t-2} and
introduce \emph{reduced} immittances $\tilde\Xi_{\alpha} \equiv \Xi_{\alpha}/\Xi_1$ ($\alpha = 3$, $\mathrm{min}$,
$\mathrm{max}$), obtaining
\begin{equation}
  \label{eq:ar:constraints-on-eta-t-3}
  \tXimin^2 \leq 
  \frac{-\RE \tilde\Xi_3 + |{\tilde\Xi_3}|^2}{\RE \tilde\Xi_3 - 1}
  \leq \tXimax^2 \, .
\end{equation}
This expression can be rewritten as
\begin{equation}
  \label{eq:ar:constraints-on-eta-t-3-geometrical}
  \begin{array}{c}
\displaystyle\Biggl[
    \biggl(\RE\tilde\Xi_3 - 
    \frac{1 + \tXimin^2}{2}\biggr)^2 +
    (\IM\tilde\Xi_3)^2 - 
    \biggl(\frac{1 - \tXimin^2}{2}\biggr)^2 \Biggr] \\[4mm]
     \times(\RE\tilde\Xi_3 - 1) \geq 0 \, , \\[4mm]
    \displaystyle\Biggl[
    \biggl(\RE\tilde\Xi_3 - 
    \frac{1 + \tXimax^2}{2}\biggr)^2 +
    (\IM\tilde\Xi_3)^2 - 
    \biggl(\frac{1 - \tXimax^2}{2}\biggr)^2 \Biggr] \\[4mm]
    \times (\RE\tilde\Xi_3 - 1) \leq 0 \, .
  \end{array}
\end{equation}
It follows that the constraints
\eqref{eq:ar:constraints-on-n} {are} equivalent to {the set of conditions written out in eq.\ \eqref{eq:ar:constraints-on-eta-t-3-set-theory}.}

\section{Numerical methods}
\label{appendixB}
A word about the different methods used for the numerical simulations presented in this
  article. Almost all the calculations whose results are presented in sections
  \ref{sec:ar:example-flat-lens} and \ref{sec:evanescent} have been made with the differential
  method \cite{VincentBook80,PopovAO00,PopovBook02,SmigajThesis}. The exceptions are the field maps
  shown in figs.\ \ref{fig:ar:lens:field-maps-uncoated} and \ref{fig:ar:lens:field-maps-gratings},
  produced with the finite-element method using the RF module of the COMSOL package. The numerical
  results reported on in section~\ref{sec:ar:example-self-collimation} have been obtained with the
  finite-difference frequency-domain method. Finally, the simulations of the magneto-optical PC
  discussed in section~\ref{sec:ar:example-superprism} have been made with a code based on the
  Fourier modal method \cite{MoharamJOSAA95} and implementing the factorisation rules for
  anisotropic materials derived in ref.~\onlinecite{PopovJOSAA01}.

%------------------------% Bibliography
%\bibliography{pc}

\begin{thebibliography}{50}
\expandafter\ifx\csname natexlab\endcsname\relax\def\natexlab#1{#1}\fi
\expandafter\ifx\csname bibnamefont\endcsname\relax
  \def\bibnamefont#1{#1}\fi
\expandafter\ifx\csname bibfnamefont\endcsname\relax
  \def\bibfnamefont#1{#1}\fi
\expandafter\ifx\csname citenamefont\endcsname\relax
  \def\citenamefont#1{#1}\fi
\expandafter\ifx\csname url\endcsname\relax
  \def\url#1{\texttt{#1}}\fi
\expandafter\ifx\csname urlprefix\endcsname\relax\def\urlprefix{URL }\fi
\providecommand{\bibinfo}[2]{#2}
\providecommand{\eprint}[2][]{\url{#2}}

\bibitem[{\citenamefont{{Notomi}}(2000)}]{NotomiPRB00}
\bibinfo{author}{\bibfnamefont{M.}~\bibnamefont{{Notomi}}},
  \bibinfo{journal}{Phys. Rev. B} \textbf{\bibinfo{volume}{62}},
  \bibinfo{pages}{10696} (\bibinfo{year}{2000}).

\bibitem[{\citenamefont{{Pendry}}(2000)}]{PendryPRL00}
\bibinfo{author}{\bibfnamefont{J.~B.} \bibnamefont{{Pendry}}},
  \bibinfo{journal}{Phys. Rev. Lett.} \textbf{\bibinfo{volume}{85}},
  \bibinfo{pages}{3966} (\bibinfo{year}{2000}).

\bibitem[{\citenamefont{Gralak et~al.}(2000)\citenamefont{Gralak, Enoch, and
  Tayeb}}]{GralakJOSAA00}
\bibinfo{author}{\bibfnamefont{B.}~\bibnamefont{Gralak}},
  \bibinfo{author}{\bibfnamefont{S.}~\bibnamefont{Enoch}}, \bibnamefont{and}
  \bibinfo{author}{\bibfnamefont{G.}~\bibnamefont{Tayeb}}, \bibinfo{journal}{J.
  Opt. Soc. Am. A} \textbf{\bibinfo{volume}{17}}, \bibinfo{pages}{1012}
  (\bibinfo{year}{2000}).

\bibitem[{\citenamefont{Enoch et~al.}(2003)\citenamefont{Enoch, Tayeb, and
  Gralak}}]{EnochTAP03}
\bibinfo{author}{\bibfnamefont{S.}~\bibnamefont{Enoch}},
  \bibinfo{author}{\bibfnamefont{G.}~\bibnamefont{Tayeb}}, \bibnamefont{and}
  \bibinfo{author}{\bibfnamefont{B.}~\bibnamefont{Gralak}},
  \bibinfo{journal}{IEEE Trans. Antennas Propag.}
  \textbf{\bibinfo{volume}{51}}, \bibinfo{pages}{2659} (\bibinfo{year}{2003}).

\bibitem[{\citenamefont{Gralak et~al.}(2006)\citenamefont{Gralak, Enoch, and
  Tayeb}}]{GralakBook06}
\bibinfo{author}{\bibfnamefont{B.}~\bibnamefont{Gralak}},
  \bibinfo{author}{\bibfnamefont{S.}~\bibnamefont{Enoch}}, \bibnamefont{and}
  \bibinfo{author}{\bibfnamefont{G.}~\bibnamefont{Tayeb}}, in
  \emph{\bibinfo{booktitle}{Metamaterials: physics and engineering
  explorations}}, edited by
  \bibinfo{editor}{\bibfnamefont{N.}~\bibnamefont{Engheta}} \bibnamefont{and}
  \bibinfo{editor}{\bibfnamefont{R.~W.} \bibnamefont{Ziolkowski}}
  (\bibinfo{publisher}{Wiley}, \bibinfo{address}{New York},
  \bibinfo{year}{2006}), chap.~\bibinfo{chapter}{10}, pp.
  \bibinfo{pages}{261--283}.

\bibitem[{\citenamefont{{Wang} and {Fan}}(2005)}]{WangAPB05}
\bibinfo{author}{\bibfnamefont{Z.}~\bibnamefont{{Wang}}} \bibnamefont{and}
  \bibinfo{author}{\bibfnamefont{S.}~\bibnamefont{{Fan}}},
  \bibinfo{journal}{Appl. Phys. B} \textbf{\bibinfo{volume}{81}},
  \bibinfo{pages}{369} (\bibinfo{year}{2005}).

\bibitem[{\citenamefont{\'{S}migaj et~al.}(2010)\citenamefont{\'{S}migaj,
  Romero-Vivas, Gralak, Magdenko, Dagens, and Vanwolleghem}}]{SmigajOL10}
\bibinfo{author}{\bibfnamefont{W.}~\bibnamefont{\'{S}migaj}},
  \bibinfo{author}{\bibfnamefont{J.}~\bibnamefont{Romero-Vivas}},
  \bibinfo{author}{\bibfnamefont{B.}~\bibnamefont{Gralak}},
  \bibinfo{author}{\bibfnamefont{L.}~\bibnamefont{Magdenko}},
  \bibinfo{author}{\bibfnamefont{B.}~\bibnamefont{Dagens}}, \bibnamefont{and}
  \bibinfo{author}{\bibfnamefont{M.}~\bibnamefont{Vanwolleghem}},
  \bibinfo{journal}{Opt. Lett.} \textbf{\bibinfo{volume}{35}},
  \bibinfo{pages}{568} (\bibinfo{year}{2010}).

\bibitem[{\citenamefont{{Dobrowolski} et~al.}(2002)\citenamefont{{Dobrowolski},
  {Poitras}, {Ma}, {Vakil}, and {Acree}}}]{DobrowolskiAO02}
\bibinfo{author}{\bibfnamefont{J.~A.} \bibnamefont{{Dobrowolski}}},
  \bibinfo{author}{\bibfnamefont{D.}~\bibnamefont{{Poitras}}},
  \bibinfo{author}{\bibfnamefont{P.}~\bibnamefont{{Ma}}},
  \bibinfo{author}{\bibfnamefont{H.}~\bibnamefont{{Vakil}}}, \bibnamefont{and}
  \bibinfo{author}{\bibfnamefont{M.}~\bibnamefont{{Acree}}},
  \bibinfo{journal}{Appl. Opt.} \textbf{\bibinfo{volume}{41}},
  \bibinfo{pages}{3075} (\bibinfo{year}{2002}).

\bibitem[{\citenamefont{{Grann} and {Moharam}}(1996)}]{GrannJOSAA96}
\bibinfo{author}{\bibfnamefont{E.~B.} \bibnamefont{{Grann}}} \bibnamefont{and}
  \bibinfo{author}{\bibfnamefont{M.~G.} \bibnamefont{{Moharam}}},
  \bibinfo{journal}{J. Opt. Soc. Am. A} \textbf{\bibinfo{volume}{13}},
  \bibinfo{pages}{988} (\bibinfo{year}{1996}).

\bibitem[{\citenamefont{Macleod}(2001)}]{MacleodBook01}
\bibinfo{author}{\bibfnamefont{H.~A.} \bibnamefont{Macleod}},
  \emph{\bibinfo{title}{Thin-film optical filters}}
  (\bibinfo{publisher}{Institute of Physics}, \bibinfo{address}{Bristol},
  \bibinfo{year}{2001}).

\bibitem[{\citenamefont{Orfanidis}(2008)}]{OrfanidisBook08}
\bibinfo{author}{\bibfnamefont{S.~J.} \bibnamefont{Orfanidis}},
  \emph{\bibinfo{title}{Electromagnetic waves and antennas}},
  \bibinfo{howpublished}{http://www.ece.rutgers.edu/\texttildelow orfanidi/ewa}
  (\bibinfo{year}{2008}).

\bibitem[{\citenamefont{{Br{\"a}uer} and {Bryngdahl}}(1994)}]{BrauerAO94}
\bibinfo{author}{\bibfnamefont{R.}~\bibnamefont{{Br{\"a}uer}}}
  \bibnamefont{and}
  \bibinfo{author}{\bibfnamefont{O.}~\bibnamefont{{Bryngdahl}}},
  \bibinfo{journal}{Appl. Opt.} \textbf{\bibinfo{volume}{33}},
  \bibinfo{pages}{7875} (\bibinfo{year}{1994}).

\bibitem[{\citenamefont{{Raguin} and {Morris}}(1993)}]{RaguinAO93March}
\bibinfo{author}{\bibfnamefont{D.~H.} \bibnamefont{{Raguin}}} \bibnamefont{and}
  \bibinfo{author}{\bibfnamefont{G.~M.} \bibnamefont{{Morris}}},
  \bibinfo{journal}{Appl. Opt.} \textbf{\bibinfo{volume}{32}},
  \bibinfo{pages}{1154} (\bibinfo{year}{1993}).

\bibitem[{\citenamefont{Raguin and Morris}(1993)}]{RaguinAO93May}
\bibinfo{author}{\bibfnamefont{D.~H.} \bibnamefont{Raguin}} \bibnamefont{and}
  \bibinfo{author}{\bibfnamefont{G.~M.} \bibnamefont{Morris}},
  \bibinfo{journal}{Appl. Opt.} \textbf{\bibinfo{volume}{32}},
  \bibinfo{pages}{2582} (\bibinfo{year}{1993}).

\bibitem[{\citenamefont{Kikuta et~al.}(2003)\citenamefont{Kikuta, Toyota, and
  Yu}}]{KikutaOR03}
\bibinfo{author}{\bibfnamefont{H.}~\bibnamefont{Kikuta}},
  \bibinfo{author}{\bibfnamefont{H.}~\bibnamefont{Toyota}}, \bibnamefont{and}
  \bibinfo{author}{\bibfnamefont{W.}~\bibnamefont{Yu}}, \bibinfo{journal}{Opt.
  Rev.} \textbf{\bibinfo{volume}{10}}, \bibinfo{pages}{63}
  (\bibinfo{year}{2003}).

\bibitem[{\citenamefont{{Xiao} et~al.}(2004)\citenamefont{{Xiao}, {Qiu},
  {Ruan}, and {He}}}]{XiaoAPL04}
\bibinfo{author}{\bibfnamefont{S.}~\bibnamefont{{Xiao}}},
  \bibinfo{author}{\bibfnamefont{M.}~\bibnamefont{{Qiu}}},
  \bibinfo{author}{\bibfnamefont{Z.}~\bibnamefont{{Ruan}}}, \bibnamefont{and}
  \bibinfo{author}{\bibfnamefont{S.}~\bibnamefont{{He}}},
  \bibinfo{journal}{Appl. Phys. Lett.} \textbf{\bibinfo{volume}{85}},
  \bibinfo{pages}{4269} (\bibinfo{year}{2004}).

\bibitem[{\citenamefont{{Botten} et~al.}(2006)\citenamefont{{Botten}, {White},
  {de Sterke}, and {McPhedran}}}]{BottenPRE06}
\bibinfo{author}{\bibfnamefont{L.~C.} \bibnamefont{{Botten}}},
  \bibinfo{author}{\bibfnamefont{T.~P.} \bibnamefont{{White}}},
  \bibinfo{author}{\bibfnamefont{C.~M.} \bibnamefont{{de Sterke}}},
  \bibnamefont{and} \bibinfo{author}{\bibfnamefont{R.~C.}
  \bibnamefont{{McPhedran}}}, \bibinfo{journal}{Phys. Rev. E}
  \textbf{\bibinfo{volume}{74}}, \bibinfo{eid}{026603} (\bibinfo{year}{2006}).

\bibitem[{\citenamefont{{Li} et~al.}(2007)\citenamefont{{Li}, {Ozbay}, {Chen},
  {Chen}, {Yang}, and {Zheng}}}]{LiJPD07}
\bibinfo{author}{\bibfnamefont{Z.}~\bibnamefont{{Li}}},
  \bibinfo{author}{\bibfnamefont{E.}~\bibnamefont{{Ozbay}}},
  \bibinfo{author}{\bibfnamefont{H.}~\bibnamefont{{Chen}}},
  \bibinfo{author}{\bibfnamefont{J.}~\bibnamefont{{Chen}}},
  \bibinfo{author}{\bibfnamefont{F.}~\bibnamefont{{Yang}}}, \bibnamefont{and}
  \bibinfo{author}{\bibfnamefont{H.}~\bibnamefont{{Zheng}}},
  \bibinfo{journal}{J. Phys. D} \textbf{\bibinfo{volume}{40}},
  \bibinfo{pages}{5873} (\bibinfo{year}{2007}).

\bibitem[{\citenamefont{{Lee} et~al.}(2008)\citenamefont{{Lee}, {Choi}, {Kim},
  {Park}, and {Kee}}}]{LeeOE08}
\bibinfo{author}{\bibfnamefont{S.-G.} \bibnamefont{{Lee}}},
  \bibinfo{author}{\bibfnamefont{J.-S.} \bibnamefont{{Choi}}},
  \bibinfo{author}{\bibfnamefont{J.-E.} \bibnamefont{{Kim}}},
  \bibinfo{author}{\bibfnamefont{H.-Y.} \bibnamefont{{Park}}},
  \bibnamefont{and} \bibinfo{author}{\bibfnamefont{C.-S.} \bibnamefont{{Kee}}},
  \bibinfo{journal}{Opt. Express} \textbf{\bibinfo{volume}{16}},
  \bibinfo{pages}{4270} (\bibinfo{year}{2008}).

\bibitem[{\citenamefont{Park et~al.}(2009)\citenamefont{Park, Lee, Park, Kim,
  and Lee}}]{ParkJOSAB09}
\bibinfo{author}{\bibfnamefont{J.-M.} \bibnamefont{Park}},
  \bibinfo{author}{\bibfnamefont{S.-G.} \bibnamefont{Lee}},
  \bibinfo{author}{\bibfnamefont{H.~Y.} \bibnamefont{Park}},
  \bibinfo{author}{\bibfnamefont{J.-E.} \bibnamefont{Kim}}, \bibnamefont{and}
  \bibinfo{author}{\bibfnamefont{M.-H.} \bibnamefont{Lee}},
  \bibinfo{journal}{J. Opt. Soc. Am. B} \textbf{\bibinfo{volume}{26}},
  \bibinfo{pages}{1967} (\bibinfo{year}{2009}).

\bibitem[{\citenamefont{{Jin} and {He}}(2007)}]{JinPLA07}
\bibinfo{author}{\bibfnamefont{Y.}~\bibnamefont{{Jin}}} \bibnamefont{and}
  \bibinfo{author}{\bibfnamefont{S.}~\bibnamefont{{He}}},
  \bibinfo{journal}{Phys. Lett. A} \textbf{\bibinfo{volume}{360}},
  \bibinfo{pages}{461} (\bibinfo{year}{2007}).

\bibitem[{\citenamefont{{Zhang} and {Li}}(2007)}]{ZhangEPJD07}
\bibinfo{author}{\bibfnamefont{B.}~\bibnamefont{{Zhang}}} \bibnamefont{and}
  \bibinfo{author}{\bibfnamefont{M.~Y.} \bibnamefont{{Li}}},
  \bibinfo{journal}{Eur. Phys. J. D} \textbf{\bibinfo{volume}{45}},
  \bibinfo{pages}{321} (\bibinfo{year}{2007}).

\bibitem[{\citenamefont{{Baba} and {Ohsaki}}(2001)}]{BabaJJAP01}
\bibinfo{author}{\bibfnamefont{T.}~\bibnamefont{{Baba}}} \bibnamefont{and}
  \bibinfo{author}{\bibfnamefont{D.}~\bibnamefont{{Ohsaki}}},
  \bibinfo{journal}{Jpn. J. Appl. Phys.} \textbf{\bibinfo{volume}{40}},
  \bibinfo{pages}{5920} (\bibinfo{year}{2001}).

\bibitem[{\citenamefont{{Baba} et~al.}(2004)\citenamefont{{Baba}, {Matsumoto},
  and {Echizen}}}]{BabaOE04}
\bibinfo{author}{\bibfnamefont{T.}~\bibnamefont{{Baba}}},
  \bibinfo{author}{\bibfnamefont{T.}~\bibnamefont{{Matsumoto}}},
  \bibnamefont{and}
  \bibinfo{author}{\bibfnamefont{M.}~\bibnamefont{{Echizen}}},
  \bibinfo{journal}{Opt. Express} \textbf{\bibinfo{volume}{12}},
  \bibinfo{pages}{4608} (\bibinfo{year}{2004}).

\bibitem[{\citenamefont{{Matsumoto} et~al.}(2005)\citenamefont{{Matsumoto},
  {Fujita}, and {Baba}}}]{MatsumotoOE05}
\bibinfo{author}{\bibfnamefont{T.}~\bibnamefont{{Matsumoto}}},
  \bibinfo{author}{\bibfnamefont{S.}~\bibnamefont{{Fujita}}}, \bibnamefont{and}
  \bibinfo{author}{\bibfnamefont{T.}~\bibnamefont{{Baba}}},
  \bibinfo{journal}{Opt. Express} \textbf{\bibinfo{volume}{13}},
  \bibinfo{pages}{10768} (\bibinfo{year}{2005}).

\bibitem[{\citenamefont{{Matsumoto} et~al.}(2006)\citenamefont{{Matsumoto},
  {Eom}, and {Baba}}}]{MatsumotoOL06}
\bibinfo{author}{\bibfnamefont{T.}~\bibnamefont{{Matsumoto}}},
  \bibinfo{author}{\bibfnamefont{K.}~\bibnamefont{{Eom}}}, \bibnamefont{and}
  \bibinfo{author}{\bibfnamefont{T.}~\bibnamefont{{Baba}}},
  \bibinfo{journal}{Opt. Lett.} \textbf{\bibinfo{volume}{31}},
  \bibinfo{pages}{2786} (\bibinfo{year}{2006}).

\bibitem[{\citenamefont{{Lawrence} et~al.}(2008)\citenamefont{{Lawrence},
  {Botten}, {Dossou}, and {de Sterke}}}]{LawrenceAPL08}
\bibinfo{author}{\bibfnamefont{F.~J.} \bibnamefont{{Lawrence}}},
  \bibinfo{author}{\bibfnamefont{L.~C.} \bibnamefont{{Botten}}},
  \bibinfo{author}{\bibfnamefont{K.~B.} \bibnamefont{{Dossou}}},
  \bibnamefont{and} \bibinfo{author}{\bibfnamefont{C.~M.} \bibnamefont{{de
  Sterke}}}, \bibinfo{journal}{Appl. Phys. Lett.}
  \textbf{\bibinfo{volume}{93}}, \bibinfo{eid}{121114} (\bibinfo{year}{2008}).

\bibitem[{\citenamefont{{Lawrence} et~al.}(2009)\citenamefont{{Lawrence},
  {Botten}, {Dossou}, {de Sterke}, and {McPhedran}}}]{LawrencePRA09}
\bibinfo{author}{\bibfnamefont{F.~J.} \bibnamefont{{Lawrence}}},
  \bibinfo{author}{\bibfnamefont{L.~C.} \bibnamefont{{Botten}}},
  \bibinfo{author}{\bibfnamefont{K.~B.} \bibnamefont{{Dossou}}},
  \bibinfo{author}{\bibfnamefont{C.~M.} \bibnamefont{{de Sterke}}},
  \bibnamefont{and} \bibinfo{author}{\bibfnamefont{R.~C.}
  \bibnamefont{{McPhedran}}}, \bibinfo{journal}{Phys. Rev. A}
  \textbf{\bibinfo{volume}{80}}, \bibinfo{eid}{023826} (\bibinfo{year}{2009}).

\bibitem[{\citenamefont{{Witzens} et~al.}(2004)\citenamefont{{Witzens},
  {Hochberg}, {Baehr-Jones}, and {Scherer}}}]{WitzensPRE04}
\bibinfo{author}{\bibfnamefont{J.}~\bibnamefont{{Witzens}}},
  \bibinfo{author}{\bibfnamefont{M.}~\bibnamefont{{Hochberg}}},
  \bibinfo{author}{\bibfnamefont{T.}~\bibnamefont{{Baehr-Jones}}},
  \bibnamefont{and}
  \bibinfo{author}{\bibfnamefont{A.}~\bibnamefont{{Scherer}}},
  \bibinfo{journal}{Phys. Rev. E} \textbf{\bibinfo{volume}{69}},
  \bibinfo{pages}{046609} (\bibinfo{year}{2004}).

\bibitem[{\citenamefont{{Momeni} and {Adibi}}(2005)}]{MomeniAPL05}
\bibinfo{author}{\bibfnamefont{B.}~\bibnamefont{{Momeni}}} \bibnamefont{and}
  \bibinfo{author}{\bibfnamefont{A.}~\bibnamefont{{Adibi}}},
  \bibinfo{journal}{Appl. Phys. Lett.} \textbf{\bibinfo{volume}{87}},
  \bibinfo{eid}{171104} (\bibinfo{year}{2005}).

\bibitem[{\citenamefont{Press et~al.}(1992)\citenamefont{Press, Flannery,
  Teukolsky, and Vetterling}}]{NumericalRecipesInC}
\bibinfo{author}{\bibfnamefont{W.~H.} \bibnamefont{Press}},
  \bibinfo{author}{\bibfnamefont{B.~P.} \bibnamefont{Flannery}},
  \bibinfo{author}{\bibfnamefont{S.~A.} \bibnamefont{Teukolsky}},
  \bibnamefont{and} \bibinfo{author}{\bibfnamefont{W.~T.}
  \bibnamefont{Vetterling}}, \emph{\bibinfo{title}{Numerical recipes in {C}}}
  (\bibinfo{publisher}{Cambridge University Press}, \bibinfo{year}{1992}).

\bibitem[{\citenamefont{{Fabre} et~al.}(2008)\citenamefont{{Fabre}, {Lalouat},
  {Cluzel}, {M{\'e}lique}, {Lippens}, {de Fornel}, and
  {Vanb{\'e}sien}}}]{FabrePRL08}
\bibinfo{author}{\bibfnamefont{N.}~\bibnamefont{{Fabre}}},
  \bibinfo{author}{\bibfnamefont{L.}~\bibnamefont{{Lalouat}}},
  \bibinfo{author}{\bibfnamefont{B.}~\bibnamefont{{Cluzel}}},
  \bibinfo{author}{\bibfnamefont{X.}~\bibnamefont{{M{\'e}lique}}},
  \bibinfo{author}{\bibfnamefont{D.}~\bibnamefont{{Lippens}}},
  \bibinfo{author}{\bibfnamefont{F.}~\bibnamefont{{de Fornel}}},
  \bibnamefont{and}
  \bibinfo{author}{\bibfnamefont{O.}~\bibnamefont{{Vanb{\'e}sien}}},
  \bibinfo{journal}{Phys. Rev. Lett.} \textbf{\bibinfo{volume}{101}},
  \bibinfo{eid}{073901} (\bibinfo{year}{2008}).

\bibitem[{\citenamefont{Vincent}(1980)}]{VincentBook80}
\bibinfo{author}{\bibfnamefont{P.}~\bibnamefont{Vincent}}, in
  \emph{\bibinfo{booktitle}{Electromagnetic theory of gratings}}, edited by
  \bibinfo{editor}{\bibfnamefont{R.}~\bibnamefont{Petit}}
  (\bibinfo{publisher}{Springer}, \bibinfo{address}{Berlin},
  \bibinfo{year}{1980}), chap.~\bibinfo{chapter}{4}, pp.
  \bibinfo{pages}{101--122}.

\bibitem[{\citenamefont{{Popov} and {Bozhkov}}(2000)}]{PopovAO00}
\bibinfo{author}{\bibfnamefont{E.}~\bibnamefont{{Popov}}} \bibnamefont{and}
  \bibinfo{author}{\bibfnamefont{B.}~\bibnamefont{{Bozhkov}}},
  \bibinfo{journal}{Appl. Opt.} \textbf{\bibinfo{volume}{39}},
  \bibinfo{pages}{4926} (\bibinfo{year}{2000}).

\bibitem[{\citenamefont{Nevi{\`e}re and Popov}(2002)}]{PopovBook02}
\bibinfo{author}{\bibfnamefont{M.}~\bibnamefont{Nevi{\`e}re}} \bibnamefont{and}
  \bibinfo{author}{\bibfnamefont{E.}~\bibnamefont{Popov}},
  \emph{\bibinfo{title}{Light propagation in periodic media. Differential
  theory and design}} (\bibinfo{publisher}{Marcel Dekker},
  \bibinfo{address}{New York}, \bibinfo{year}{2002}).

\bibitem[{\citenamefont{{{\'S}migaj}}(2007)}]{SmigajThesis}
\bibinfo{author}{\bibfnamefont{W.}~\bibnamefont{{{\'S}migaj}}}, Master's
  thesis, \bibinfo{school}{Faculty of Physics, Adam Mickiewicz University in
  Pozna{\'n}} (\bibinfo{year}{2007}).

\bibitem[{\citenamefont{Veselago}(1968)}]{VeselagoSPU68}
\bibinfo{author}{\bibfnamefont{V.~G.} \bibnamefont{Veselago}},
  \bibinfo{journal}{Sov. Phys. Usp.} \textbf{\bibinfo{volume}{10}},
  \bibinfo{pages}{509} (\bibinfo{year}{1968}).

\bibitem[{\citenamefont{{{\'S}migaj} and
  {Gralak}}(2008{\natexlab{a}})}]{SmigajPRB08}
\bibinfo{author}{\bibfnamefont{W.}~\bibnamefont{{{\'S}migaj}}}
  \bibnamefont{and} \bibinfo{author}{\bibfnamefont{B.}~\bibnamefont{{Gralak}}},
  \bibinfo{journal}{Phys. Rev. B} \textbf{\bibinfo{volume}{77}},
  \bibinfo{eid}{235445} (\bibinfo{year}{2008}{\natexlab{a}}).

\bibitem[{\citenamefont{{{\'S}migaj} and
  {Gralak}}(2008{\natexlab{b}})}]{SmigajSPIE08}
\bibinfo{author}{\bibfnamefont{W.}~\bibnamefont{{{\'S}migaj}}}
  \bibnamefont{and} \bibinfo{author}{\bibfnamefont{B.}~\bibnamefont{{Gralak}}},
  \bibinfo{journal}{Proc. SPIE} \textbf{\bibinfo{volume}{6987}},
  \bibinfo{eid}{698726} (\bibinfo{year}{2008}{\natexlab{b}}).

\bibitem[{\citenamefont{{{\'S}migaj}
  et~al.}(2009{\natexlab{a}})\citenamefont{{{\'S}migaj}, {Gralak}, {Pierre},
  and {Tayeb}}}]{SmigajSPP4}
\bibinfo{author}{\bibfnamefont{W.}~\bibnamefont{{{\'S}migaj}}},
  \bibinfo{author}{\bibfnamefont{B.}~\bibnamefont{{Gralak}}},
  \bibinfo{author}{\bibfnamefont{R.}~\bibnamefont{{Pierre}}}, \bibnamefont{and}
  \bibinfo{author}{\bibfnamefont{G.}~\bibnamefont{{Tayeb}}}, in
  \emph{\bibinfo{booktitle}{SPP4 Surface Plasmon Photonics Conference}}
  (\bibinfo{year}{2009}{\natexlab{a}}).

\bibitem[{\citenamefont{{{\'S}migaj}
  et~al.}(2009{\natexlab{b}})\citenamefont{{{\'S}migaj}, {Gralak}, {Pierre},
  and {Tayeb}}}]{SmigajOL09}
\bibinfo{author}{\bibfnamefont{W.}~\bibnamefont{{{\'S}migaj}}},
  \bibinfo{author}{\bibfnamefont{B.}~\bibnamefont{{Gralak}}},
  \bibinfo{author}{\bibfnamefont{R.}~\bibnamefont{{Pierre}}}, \bibnamefont{and}
  \bibinfo{author}{\bibfnamefont{G.}~\bibnamefont{{Tayeb}}},
  \bibinfo{journal}{Opt. Lett.} \textbf{\bibinfo{volume}{34}},
  \bibinfo{pages}{3532} (\bibinfo{year}{2009}{\natexlab{b}}).

\bibitem[{\citenamefont{{Kosaka} et~al.}(1998)\citenamefont{{Kosaka},
  {Kawashima}, {Tomita}, {Notomi}, {Tamamura}, {Sato}, and
  {Kawakami}}}]{KosakaPRB98}
\bibinfo{author}{\bibfnamefont{H.}~\bibnamefont{{Kosaka}}},
  \bibinfo{author}{\bibfnamefont{T.}~\bibnamefont{{Kawashima}}},
  \bibinfo{author}{\bibfnamefont{A.}~\bibnamefont{{Tomita}}},
  \bibinfo{author}{\bibfnamefont{M.}~\bibnamefont{{Notomi}}},
  \bibinfo{author}{\bibfnamefont{T.}~\bibnamefont{{Tamamura}}},
  \bibinfo{author}{\bibfnamefont{T.}~\bibnamefont{{Sato}}}, \bibnamefont{and}
  \bibinfo{author}{\bibfnamefont{S.}~\bibnamefont{{Kawakami}}},
  \bibinfo{journal}{Phys. Rev. B} \textbf{\bibinfo{volume}{58}},
  \bibinfo{pages}{R10096} (\bibinfo{year}{1998}).

\bibitem[{\citenamefont{{Prather} et~al.}(2007)\citenamefont{{Prather}, {Shi},
  {Murakowski}, {Schneider}, {Sharkawy}, {Chen}, {Miao}, and
  {Martin}}}]{PratherJPD07}
\bibinfo{author}{\bibfnamefont{D.~W.} \bibnamefont{{Prather}}},
  \bibinfo{author}{\bibfnamefont{S.}~\bibnamefont{{Shi}}},
  \bibinfo{author}{\bibfnamefont{J.}~\bibnamefont{{Murakowski}}},
  \bibinfo{author}{\bibfnamefont{G.~J.} \bibnamefont{{Schneider}}},
  \bibinfo{author}{\bibfnamefont{A.}~\bibnamefont{{Sharkawy}}},
  \bibinfo{author}{\bibfnamefont{C.}~\bibnamefont{{Chen}}},
  \bibinfo{author}{\bibfnamefont{B.~L.} \bibnamefont{{Miao}}},
  \bibnamefont{and} \bibinfo{author}{\bibfnamefont{R.}~\bibnamefont{{Martin}}},
  \bibinfo{journal}{J. Phys. D} \textbf{\bibinfo{volume}{40}},
  \bibinfo{pages}{2635} (\bibinfo{year}{2007}).

\bibitem[{\citenamefont{Vanwolleghem et~al.}(2009)\citenamefont{Vanwolleghem,
  Checoury, {\'S}migaj, Gralak, Magdenko, Postava, Dagens, Beauvillain, and
  Lourtioz}}]{VanwolleghemPRB09}
\bibinfo{author}{\bibfnamefont{M.}~\bibnamefont{Vanwolleghem}},
  \bibinfo{author}{\bibfnamefont{X.}~\bibnamefont{Checoury}},
  \bibinfo{author}{\bibfnamefont{W.}~\bibnamefont{{\'S}migaj}},
  \bibinfo{author}{\bibfnamefont{B.}~\bibnamefont{Gralak}},
  \bibinfo{author}{\bibfnamefont{L.}~\bibnamefont{Magdenko}},
  \bibinfo{author}{\bibfnamefont{K.}~\bibnamefont{Postava}},
  \bibinfo{author}{\bibfnamefont{B.}~\bibnamefont{Dagens}},
  \bibinfo{author}{\bibfnamefont{P.}~\bibnamefont{Beauvillain}},
  \bibnamefont{and} \bibinfo{author}{\bibfnamefont{J.-M.}
  \bibnamefont{Lourtioz}}, \bibinfo{journal}{Phys. Rev. B}
  \textbf{\bibinfo{volume}{80}}, \bibinfo{eid}{121102(R)}
  (\bibinfo{year}{2009}).

\bibitem[{\citenamefont{Sakoda}(2001)}]{SakodaBook01}
\bibinfo{author}{\bibfnamefont{K.}~\bibnamefont{Sakoda}},
  \emph{\bibinfo{title}{Optical properties of photonic crystals}}
  (\bibinfo{publisher}{Springer}, \bibinfo{address}{Berlin},
  \bibinfo{year}{2001}).

\bibitem[{\citenamefont{Luo et~al.}(2003)\citenamefont{Luo, Johnson,
  Joannopoulos, and Pendry}}]{LuoPRB03}
\bibinfo{author}{\bibfnamefont{C.}~\bibnamefont{Luo}},
  \bibinfo{author}{\bibfnamefont{S.~G.} \bibnamefont{Johnson}},
  \bibinfo{author}{\bibfnamefont{J.~D.} \bibnamefont{Joannopoulos}},
  \bibnamefont{and} \bibinfo{author}{\bibfnamefont{J.~B.}
  \bibnamefont{Pendry}}, \bibinfo{journal}{Phys. Rev. B}
  \textbf{\bibinfo{volume}{68}}, \bibinfo{pages}{045115}
  (\bibinfo{year}{2003}).

\bibitem[{\citenamefont{{\'S}migaj et~al.}(2011)\citenamefont{{\'S}migaj,
  Lalanne, Yang, Paul, Rockstuhl, and Lederer}}]{SmigajAPL11}
\bibinfo{author}{\bibfnamefont{W.}~\bibnamefont{{\'S}migaj}},
  \bibinfo{author}{\bibfnamefont{P.}~\bibnamefont{Lalanne}},
  \bibinfo{author}{\bibfnamefont{J.}~\bibnamefont{Yang}},
  \bibinfo{author}{\bibfnamefont{T.}~\bibnamefont{Paul}},
  \bibinfo{author}{\bibfnamefont{C.}~\bibnamefont{Rockstuhl}},
  \bibnamefont{and} \bibinfo{author}{\bibfnamefont{F.}~\bibnamefont{Lederer}},
  \bibinfo{journal}{Appl. Phys. Lett.} \textbf{\bibinfo{volume}{98}},
  \bibinfo{eid}{111107} (\bibinfo{year}{2011}).

\bibitem[{\citenamefont{{Pierre} and Gralak}(2008)}]{PierreJMO07}
\bibinfo{author}{\bibfnamefont{R.}~\bibnamefont{{Pierre}}} \bibnamefont{and}
  \bibinfo{author}{\bibfnamefont{B.}~\bibnamefont{Gralak}},
  \bibinfo{journal}{J. Mod. Opt.} \textbf{\bibinfo{volume}{55}},
  \bibinfo{pages}{1759} (\bibinfo{year}{2008}).

\bibitem[{\citenamefont{Moharam et~al.}(1995)\citenamefont{Moharam, Grann,
  Pommet, and Gaylord}}]{MoharamJOSAA95}
\bibinfo{author}{\bibfnamefont{M.~G.} \bibnamefont{Moharam}},
  \bibinfo{author}{\bibfnamefont{E.~B.} \bibnamefont{Grann}},
  \bibinfo{author}{\bibfnamefont{D.~A.} \bibnamefont{Pommet}},
  \bibnamefont{and} \bibinfo{author}{\bibfnamefont{T.~K.}
  \bibnamefont{Gaylord}}, \bibinfo{journal}{J. Opt. Soc. Am. A}
  \textbf{\bibinfo{volume}{12}}, \bibinfo{pages}{1068} (\bibinfo{year}{1995}).

\bibitem[{\citenamefont{{Popov} and {Nevi{\`e}re}}(2001)}]{PopovJOSAA01}
\bibinfo{author}{\bibfnamefont{E.}~\bibnamefont{{Popov}}} \bibnamefont{and}
  \bibinfo{author}{\bibfnamefont{M.}~\bibnamefont{{Nevi{\`e}re}}},
  \bibinfo{journal}{J. Opt. Soc. Am. A} \textbf{\bibinfo{volume}{18}},
  \bibinfo{pages}{2886} (\bibinfo{year}{2001}).

\end{thebibliography}
%\input{ARG.bbl}

%------------------------% Bibliography

\end{document}